\documentclass[aps,pra,twocolumn,groupedaddress,showpacs]{revtex4-1}
\usepackage{graphicx}
\usepackage{mathrsfs,amsmath,amssymb} 
\usepackage{lipsum}
\usepackage[usenames]{color}
\usepackage{amsmath}

\newcommand{\beq}{\begin{equation}}
\newcommand{\eeq}{\end{equation}}
\newcommand{\bea}{\begin{eqnarray}}
\newcommand{\eea}{\end{eqnarray}}
\newcommand{\bra}[1]{\langle #1|}
\newcommand{\ket}[1]{|#1\rangle}
\newcommand{\braket}[2]{\langle #1|#2\rangle}

\newcommand{\bket}[1]{\langle #1 \rangle}

\begin{document}


\title{Quantum Correlations in Optical Metrology: \break Heisenberg-limited Phase Estimation without mode Entanglement}


\author{Jaspreet Sahota}
\email[]{jsahota@physics.utoronto.ca}

\author{Nicol\'{a}s Quesada}

\affiliation{CQIQC and IOS, Department of Physics, University of Toronto, 60 Saint George Street, Toronto, Ontario M5S 1A7, Canada}


\date{\today}

\begin{abstract}
The quantum fisher information and quantum correlation parameters are employed to study the application of non-classical light to the problem of parameter estimation. It is shown that the optimal measurement sensitivity of a quantum state is determined by its inter-mode correlations (which depends of path-entanglement) and intra-mode correlations (which depends of the photon statistics). In light of these results, we consider the performance of quantum-enhanced optical interferometers.  Furthermore, we propose a Heisenberg-limited metrology protocol involving standard elements from passive and active linear optics, for which the quantum Cram\'{e}r-Rao bound is saturated with an intensity measurement. Interestingly, the quantum advantage for this scheme is derived solely from the non-classical photon statistics of the probe state and does not depend of entanglement. We study the performance of this scheme in the presence of realistic losses and consequently predict a substantial enhancement over the shot-noise limit with current technological capabilities.
\end{abstract}



\maketitle


\section{Introduction}

The requirement of performing extremely sensitive and high-resolution measurements is ubiquitous in the fundamental and applied sciences. Some examples of this include: gravitational wave detection via optical interferometers \cite{LIGO11}, Ramsey interferometry for measuring properties of atoms and molecules \cite{RevModPhys.81.1051}, and nano-device fabrication using optical lithography \cite{Pavel2013259}. The fundamental limits on measurement sensitivity and resolution are ultimately dictated by the laws of quantum mechanics. The field of quantum metrology is concerned with determining these limits and providing protocols to realize them by exploiting quantum resources \cite{Giovannetti:2011ys, giovannetti04}.

\begin{figure}[b]
\centering
\includegraphics[width=7.7 cm]{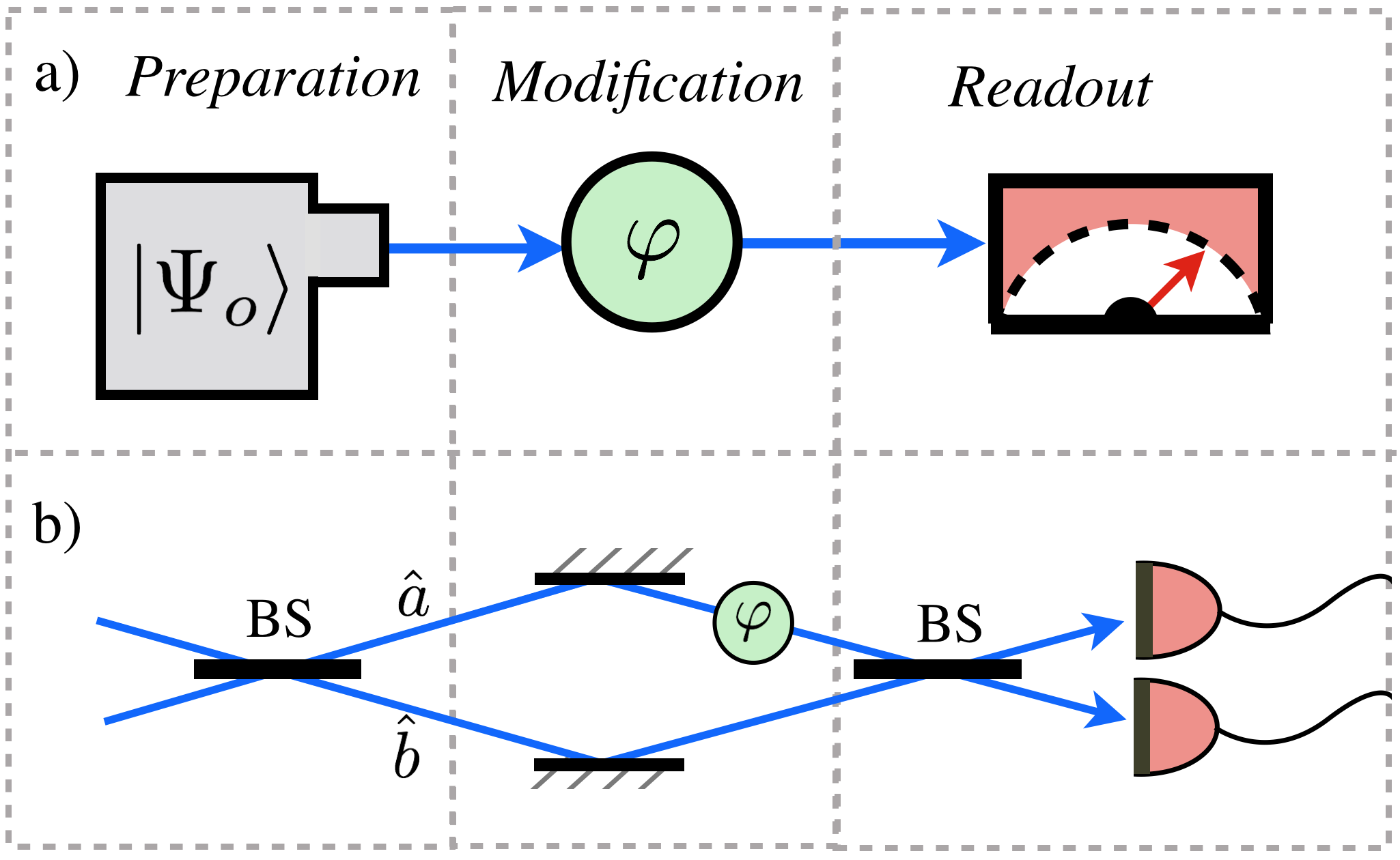}
\caption{(Color online)  (a) The three steps of a metrology protocol: Probe preparation, probe modification, and probe readout. (b) Example: the MZI, where a relative phase $\varphi$ between modes $\hat{a}$ and $\hat{b}$ is inferred by observing the photon intensities of the output beams of the interferometer.}
\label{fig:scheme}
\end{figure}

We can divide any measurement protocol into three stages: probe preparation, probe modification, and probe readout (see Fig. \ref{fig:scheme}a). Initially, a quantum state $\ket{\Psi_o}$ is prepared, which serves as the probing state for the measurement \footnote{Note that this can be generalized to mixed states, however we will mainly be concerned with pure states in this article.}. Then, $\ket{\Psi_o}$ is modified by some physical mechanism, resulting in a state $\ket{\Psi_\varphi}$, which depends on a parameter $\varphi$ that we are interested in estimating. Finally, we measure the expectation value of an observable $\hat{O}$ corresponding to $\ket{\Psi_\varphi}$ and the resulting signal $\mathcal{S}(\varphi)= \bra{\Psi_\varphi}\hat{O}\ket{\Psi_\varphi}$ is used to estimate $\varphi$.  As a concrete example, consider a coherent laser field $\ket{\alpha}$ that is fed into a 50:50 beam splitter (BS) to obtain the output probe state $\hat V\ket{\alpha}\otimes \ket{0}=\ket{\frac{i \alpha}{\sqrt{2}}}\otimes \ket{\frac{\alpha}{\sqrt{2}}}$ where $\hat V=e^{i \frac{\pi}{4}(\hat a^\dagger \otimes \hat b+ \hat a \otimes \hat b^\dagger)}$ represents the operation of a 50:50 BS. The two spatial modes of this state, corresponding to the arms inside a Mach-Zehnder interferometer (MZI), acquire a relative phase $\varphi$ as described by  $\hat{U}(\varphi)= {\exp[ \frac{i \varphi}{2} (\hat{a}^{\dag} \hat{a} - \hat{b}^{\dag} \hat{b}) ]}$. Finally, the resulting state is passed through a second 50:50 BS and a relative intensity measurement $\langle \hat{O} \rangle = \langle \hat{a}^{\dag} \hat{a}-\hat{b}^{\dag} \hat{b} \rangle$ is performed on the output beams. This measurement signal is used to estimate $\varphi$ with estimation error $\Delta \varphi$ that is bounded from below by the shot-noise limit (SNL): $\Delta \varphi \ge 1/\sqrt{\bar{n}}$, where $\bar{n}= |\alpha|^2$ is the mean photon number of $\ket{\alpha}$  (see Fig. \ref{fig:scheme}b for a diagram of a MZI). Specifically speaking, one should take into account the number of times the experiment is carried out. If $\Delta \varphi$ is determined from $k$ experiments, each of which employ a probe state containing $\bar{n}$ photons on average, then the SNL is given by $\Delta \varphi = 1/\sqrt{k \bar{n}}$. However, one typically omits this $k$ term for the sake of brevity as we will also do in the following.

In 1981, Caves showed that the unused port of the laser-MZI (as described above) can be injected with a squeezed vacuum state to attain sub-SNL phase error \cite{Caves81}. The probe state created by such a preparation procedure is path-entangled inside the MZI. The success of this protocol initiated efforts to exploit the quantum nature of light in order to reach the fundamental limits of quantum metrology. In another approach, the maximally path-entangled NOON state, $\ket{\text{NOON}} = (\ket{n,0}+\ket{0,n})/\sqrt{2}$, is created inside the MZI. This is the optimal $n$-photon state for noiseless quantum interferometry as it attains a phase error $\Delta \varphi = 1/n$ (this is called this Heisenberg limit)  \cite{PhysRevA.61.043811, PhysRevLett.85.2733, PhysRevA.66.013804, doi:10.1080/0950034021000011536, PhysRevLett.75.2944,PhysRevLett.77.2352, PhysRevLett.96.010401, PhysRevLett.105.180402, PhysRevA.85.041802, PhysRevLett.108.210404}.  Much work has focused on developing quantum schemes that can realize this limit \cite{PhysRevLett.56.1515, PhysRevA.33.4033,PhysRevLett.71.1355,Jacobson:1995fk,PhysRevA.56.R1083, PhysRevLett.100.073601, PhysRevLett.110.163604,PhysRevA.61.043811, PhysRevA.66.013804, doi:10.1080/0950034021000011536}. Unfortunately, these states are easily decohered in the presence of losses, consequently losing their sensitivity to phase changes \cite{Gilbert:08, PhysRevA.80.063803}. This has initiated efforts to find $n$-photon probe states that are optimal in the presence losses \cite{PhysRevLett.102.040403, PhysRevA.83.021804, Escher:2011zr, escher2011quantum, PhysRevA.78.063828}.  Aside from improving sensitivity, another challenge of quantum metrology is to generate bright probe states (i.e. states with a large mean photon number) that yield high resolution measurement. Ref.\cite{PhysRevLett.110.163604} and Ref.\cite{sahota13} propose different methods of creating bright entangled states inside a MZI. 

Most quantum technologies rely on the resources of quantum entanglement to accomplish tasks deemed impossible by classical physics. Some examples of such technologies includes quantum computing \cite{nielsen2010quantum}, quantum teleportation \cite{PhysRevLett.70.1895}, quantum cryptography \cite{RevModPhys.74.145}, quantum-enhanced photodetector calibration \cite{Migdall:02}, quantum imaging \cite{doi:10.1080/09500340600589382}, and lidar \cite{Lloyd12092008}. In addition, the aforementioned metrology protocols employ probe states that are path-entangled inside the MZI to beat the classical limit (i.e. the SNL). This has fostered the view that entanglement is necessary for quantum-enhanced metrology. Refs.\cite{PhysRevA.81.022108, Datta2012} have challenged this view by demonstrating that the advantage obtained via entanglement is contingent on the measurement employed and the way in which the unknown parameter is imprinted in the probe state, and have further shown that under certain conditions entanglement can even be disadvantageous in a metrology protocol. 

\begin{table*}
 \begin{tabular}{lccc}
 \toprule
  \bf{Probe State}&$\mathcal{Q}$&$\mathcal{J}$&$\mathfrak{F}$ \\
 \toprule
 $\ket{\frac{i \alpha}{\sqrt{2}}}\otimes \ket{\frac{\alpha}{\sqrt{2}}}$  &$0$&$0$& $\bar{n}$\\
  \colrule
  NOON State & $\frac{\bar{n}}{2}-1$& $-1$& $\bar{n}^2$\\
    \colrule
$\ket{s(r,0)}\otimes\ket{s(r,0)}$&$\bar{n}+1$& 0 & $\bar{n}^2+2 \bar{n}$\\
 \colrule
 Caves State & $\frac{1+2\bar{n}+\sqrt{\bar{n}(\bar{n}+2)}}{4}$ & $\frac{1-\sqrt{\bar{n}(\bar{n}+2)}}{5+ 2\bar{n}+\sqrt{\bar{n}(\bar{n}+2)}}$ & $\frac{2\bar{n}+\bar{n}\sqrt{\bar{n}(\bar{n}+2)}+ \bar{n}^2}{2}$ \\
 \colrule
Amplified Bell State &$\frac{5\bar{n}-{11}/{\bar{n}}+2}{8}$& $\frac{-(\bar{n}+1)^2}{5\bar{n}^2+10\bar{n}-11}$ & $\frac{(3\bar{n}^2+6\bar{n}-5)}{4}$\\
\colrule
Twin Fock State & $\frac{\left(\bar n/2 -1\right)}{2}$ &$-1$ & $\frac{\bar n^2}{2}+\bar{n}$ \\
\colrule
Two Mode Squeezed Vacuum & $\bar{n}$ & $1$ & $0$ \\
\colrule
Entangled Coherent State & $\frac{\bar{n}}{2}$ & $\frac{-1}{1+\frac{2}{\bar{n}}}$ & $\bar{n}^2+\bar{n}$\\
\colrule
 \end{tabular}
 \caption{The Mandel parameter $\mathcal{Q}$, mode correlation parameter $\mathcal{J}$, and the quantum Fisher information $\mathfrak{F}$ are listed for various MZI path-symmetric probe states: Laser light, NOON states \cite{PhysRevLett.85.2733}, twin squeezed vacuums, Caves state \cite{Caves81} (assuming the laser intensity equals the squeezed state intensity), the amplified bell state \cite{sahota13}, the twin Fock state \cite{PhysRevLett.71.1355}, two mode squeezed vacuum and the entangled coherent state\cite{PhysRevLett.107.083601} (Assuming the mean number of photons $\bar{n}$ is such that $e^{-\bar{n}} \ll 1$). Note that $\mathcal{J}$ is always between -1 and 1, meaning  that the super-sensitivity of these probe states results from the scaling of $\mathcal{Q}$ rather than  mode entanglement when $\bar{n}$ is large.}
  \label{table}
\end{table*}

In this manuscript, we show that the metrological power of quantum light is determined not only by the inter-mode correlations (i.e. path-entanglement), but also by intra-mode correlations (determined by the photon statistics within each mode and quantified by the Mandel $\mathcal{Q}$-parameter). Hence, we are able to identity scenarios where the fragile property of path entanglement can be discarded, and a ``quantum-enhancement" can be attained by leveraging solely the non-classical photon statistics of quantum light. 
Finally, motivated by these findings, we are able to describe a metrology protocol which attains Heisenberg-limited phase sensitivity without the aid of entanglement. Conveniently, this proposal relies only on commonly employed technologies of passive and active linear optics. Based on the analysis provided, we predict a significant improvement over the SNL in the presence of realistic losses.

\section{Quantum Fisher information}\label{qfi}

For a general metrology strategy as shown in Fig. \ref{fig:scheme}a, the error $\Delta^2 \varphi$ in the estimated phase is bounded from below by the inverse Fisher information, as given by the Cram\'{e}r-Rao bound, $\Delta^2 \varphi \ge {1/F}$. The Fisher information $F$ is determined by the measurement statistics used to estimate $\varphi$ \cite{1976quantum}. That is, if a POVM $\{ \hat{\Pi}_i \}$ describes a measurement on the modified probe $\ket{\Psi_\varphi}$, then
\beq
F= \sum_{i} \frac{1}{\bra{\Psi_\varphi} \hat{\Pi}_i  \ket{\Psi_\varphi}} \left( \frac{\partial \bra{\Psi_\varphi} \hat{\Pi}_i  \ket{\Psi_\varphi} }{\partial \varphi} \right)^2.
\eeq
Optimizing over measurements, we obtain the quantum Fisher information, which for a pure state equals $\mathfrak{F} =4 \left( \braket{\Psi_\varphi^{\prime}}{\Psi_\varphi^{\prime}} - |\braket{\Psi_\varphi^{\prime}}{\Psi_\varphi}|^2 \right)$, where primes denote derivatives with respect to $\varphi$. Hence, the quantum Cram\'{e}r-Rao bound reads  \cite{1976quantum, Caves94,braunstein1996generalized}
\beq
\Delta^2 \varphi \ge \frac{1}{F} \ge \frac{1}{\mathfrak{F}}. \label{QCR}
\eeq 
In the case of a MZI, where $\hat{U}(\varphi)= {\exp[ \frac{i \varphi}{2} (\hat{a}^{\dag} \hat{a} - \hat{b}^{\dag} \hat{b}) ]}$ describe the incurred phase shift, we have $\mathfrak{F}={\Delta^2 (\hat{a}^{\dag} \hat{a} - \hat{b}^{\dag} \hat{b})}$ \cite{sahota13}. This expands to
\beq
\mathfrak{F}= \Delta^2 (\hat{n}_a) + \Delta^2(\hat{n}_{b}) -2 \, \text{Cov}[ \hat{n}_a,\hat{n}_b], \label{QF1}
\eeq
where $\hat{n}_a=\hat{a}^{\dag} \hat{a}$, $\hat{n}_b=\hat{b}^{\dag} \hat{b}$ and $\text{Cov}[ \hat{n}_a,\hat{n}_b]=({\langle \hat{n}_a \otimes \hat{n}_b  \rangle} - \langle \hat{n}_a \rangle \langle \hat{n}_b\rangle)$. Eq. (\ref{QF1}) conveys the resources important for quantum-enhanced optical interferometry. The first two terms describe the photon statistics in each arm of the MZI and the $\text{Cov}[ \hat{n}_a,\hat{n}_b]$ term describes correlations between the arms of the MZI. To make this more explicit, we introduce the Mandel $\mathcal{Q}$-parameter,  $\mathcal{Q}_a={(\Delta^2 \hat{n}_a-\langle \hat{n}_a\rangle})/{\langle \hat{n}_a\rangle}$ (likewise for mode $\hat{b}$), and the mode correlation parameter $\mathcal{J} = \text{Cov}[ \hat{n}_a,\hat{n}_b]/(\Delta \hat{n}_a \Delta \hat{n}_b)$ \cite{gerry2005introductory}. Therefore,
\bea
\mathfrak{F} &=& \langle \hat{n}_a \rangle (1+\mathcal{Q}_a) +\langle \hat{n}_b \rangle (1+\mathcal{Q}_b) \nonumber \\
&-&2\sqrt{\langle \hat{n}_a \rangle \langle \hat{n}_b \rangle (1+\mathcal{Q}_a) (1+\mathcal{Q}_b) }\mathcal{J}. 
\eea
The probe states of interest in most phase estimation protocols are symmetric with respect to an exchange (nonphysical) of the MZI arms \cite{PhysRevA.79.033822}. This path symmetry assumption implies,  $ \langle \hat{n}_a \rangle= \langle \hat{n}_b \rangle=\bar{n}/2$ and  $\langle \hat{n}_a^2 \rangle =  \langle \hat{n}_b^2 \rangle$. Therefore,  
\beq
\mathfrak{F}= \bar{n}\left(1+\mathcal{Q} \right)\left(1-\mathcal{J} \right), \label{QFPS}
\eeq
where $\bar{n}$ is the average number of photons in the probe state. For $\mathcal{Q}>0$ $(-1<\mathcal{Q}<0)$, the photon statistics within each arm of the MZI is super-poissonian (sub-poissonian). The inter-mode correlations (between the arms of the MZI) are described by $\mathcal{J}$, which depends on the path-entanglement of the probe state. We note that $\mathcal{J}$ ranges between -1 and 1, meaning that $\mathcal{Q}$ should scale as $\bar{n}$ in order to obtain Heisenberg error scaling when $\bar{n}$ is large. Surprisingly, this suggests that the metrological advantage of non-classical light is derived primarily from the photon statistics and not path-entanglement. This fact is depicted in Table \ref{table}, where we have tabulated the values of $\mathcal{Q}$, $\mathcal{J}$, and $\mathfrak{F}$ for various path-symmetric MZI probe states that are of interest in quantum metrology. Furthermore, it is evident from Eq. (\ref{QFPS}) that even separable state (i.e. states for which $\mathcal{J}=0$) can surpass the SNL, provided that the photon statistics is super-poissonian. For example, the twin single-mode squeezed vacuum probe state $\ket{s(r,0)}\otimes\ket{s(r,0)}$, where
\beq
\ket{s(r,\varphi)}=\sum_{j=0}^{\infty} (-1)^{j} \frac{\sqrt{(2j)!}}{2^{j}j!}  \left[ \frac{(\tanh r)^{j} }{\sqrt{\cosh r }} \right] e^{i 2 j \varphi } \ket{2 j} \label{sqv}
\eeq
is the single-mode squeezed vacuum with phase $\varphi$, yields Heisenberg-limited phase error. This follows from Eq. (\ref{QFPS}), since $\mathfrak{F}= \bar{n}(\bar{n}+2)$, where $\bar{n}=\sinh^2{r}$. This sensitivity can be attained in principle by performing a parity measurement on one of the output modes of the MZI \cite{PhysRevLett.104.103602}. Also, note that in Ref. \cite{Lang2014} $\mathfrak{F}=\bar{n}(\bar{n}+2)$ is optimal under the constraint of fixed $\bar{n}$.
In the case when the photon number is fixed (\emph{i.e.} are no number fluctuations) the optimal state to be injected into a MZI is the twin Fock state $\ket{n/2} \otimes \ket{n/2}$ giving the state $\hat V \ket{n/2} \otimes \ket{n/2}$ inside the interferometer \cite{PhysRevLett.71.1355,Lang2014}.

\section{Heisenberg-limited sensitivity without entanglement}

Now we discuss a practical metrology strategy that provides Heisenberg-limited phase error in the lossless case without employing path-entanglement (i.e. $\mathcal{J}=0$). In this scenario, a quantum advantage is obtained because the photon statistics of the probe state is super-poissonian. Finally, we investigate the performance of this scheme in the presence of realistic losses, and conclude that a significant improvement over the SNL should be observable with current technological capabilities.    

The probe state for this protocol is generated by pumping a strong coherent field $\ket{\beta}$ into a $\chi^{(2)}$ crystal. On occasion, a photon from the pump field is converted into a pair of identical photons of the probe field, each of which has half the frequency of the pump field photon. This process, known as degenerate parametric down-conversion, is described by the interaction hamiltonian
\beq
\hat{H}=i \hbar \chi^{(2)} \left(\hat{a}^2\otimes  \hat{b}^{\dag} -  \hat{a}^{\dag 2} \otimes \hat{b} \right), \label{int}
\eeq  
where the $\hat{b}$, $\hat{b}^{\dag}$ correspond to the pump field and $\hat{a}$, $\hat{a}^{\dag}$ correspond to the probe state. As a result of this interaction, the state created in mode $\hat{a}$ is $\ket{s(r,0)}$, a single-mode squeezed vacuum as described in Eq. (\ref{sqv}) \footnote{Another consequence of (\ref{int}) is that the phase of the pump field is inherited by the squeezed vacuum created in mode $\hat{a}$. Without loss of generality, we have set this phase to be 0.}. Note that in our proposal the strong classical beam prepared in mode $\hat b$ will serve as the phase reference for the phase $\varphi$ imprinted on the squeezed state. In the undepleted pump regime, the output of the $\chi^{(2)}$ crystal is a product state of the coherent classical field and the single-mode squeezed vacuum to be employed as the probe state. Therefore we need not consider the classical pump field in the calculation of the Fisher information \cite{Raf2014}.

\begin{figure}[b]
\centering
\includegraphics[width=8.5 cm]{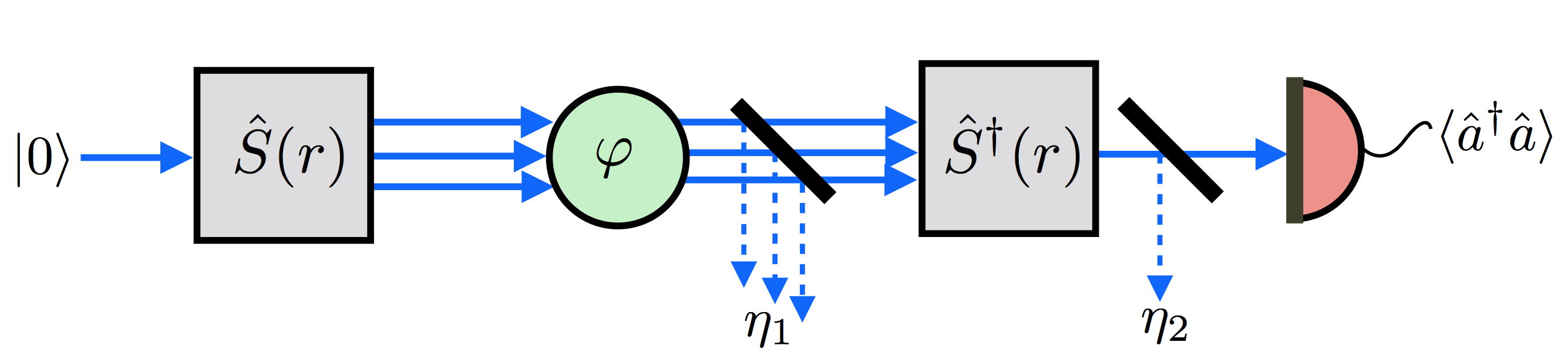}
\caption{(Color online)  The probe state for this scheme is obtained by squeezing the vacuum state $\ket{0}$ to obtain $\ket{s(r,0)}$. A phase shift occurs between $\ket{s(r,0)}$ and the pump field used to create it, which results in the modified probe state $\ket{s(r,\varphi)}$. Finally, the squeezing operation is reversed and a photon intensity measurement is performed on the resulting state. Note that the pump field has been omitted from this diagram for simplicity.}
\label{fig:scheme1}
\end{figure} 

Now we are interested in reversing this squeezing operation so that $\ket{s(r,0)}$ is anti-squeezed back to the vacuum $\ket{0}$. This can be accomplished by either retro-reflecting the output modes (both pump and down converted photons) of the $\chi^{(2)}$ crystal back onto itself, or by sending these modes into a second $\chi^{(2)}$ crystal, which serves to reverse the effect of the first crystal. Both of these techniques are feasible from an experimental standpoint \cite{PhysRevLett.72.629,NYAS:NYAS61, PhysRevLett.87.123603, PhysRevLett.88.113601, PhysRevLett.89.037904, PhysRevLett.102.020404, Hudelist:2014dz}. In order to successfully obtain $\ket{0}$ after anti-squeezing $\ket{s(r,0)}$, there must be a definite phase relation between the probe mode $\hat{a}$ and the pump mode $\hat{b}$. If mode $\hat{a}$ acquires an unknown phase $\varphi$ relative to mode $\hat{b}$, then the output of this protocol will be  
\beq
\ket{\Psi_f}= \hat{S}^{\dag}(r) \hat{U}(\varphi) \hat{S}(r)\ket{0}=\hat{S}^{\dag}(r) \hat{U}(\varphi)  \ket{s(r,0)}.
\eeq 
Here we are denoting the squeezing and anti-squeezing operations as $\hat{S}(r)=e^{\frac{r}{2}( \hat{a}^2- \hat{a}^{\dagger 2})}$ and  $\hat{S}^{\dag}(r)$ respectively. Additionally, the phase shift acquired by mode $\hat{a}$ is described by $ \hat{U}(\varphi)= e^{i \varphi \hat{a}^{\dag}\hat{a}}$, so that $\ket{\Psi_f}= \hat{S}^{\dag}(r) \ket{s(r,\varphi)}$ (see Fig. \ref{fig:scheme1}). 

The unknown phase $\varphi$ can be inferred from the photon intensity of $\ket{\Psi_f}$, which we determine to be
\bea
\mathcal{S}(\bar{n},\varphi) &=&\bra{\Psi_f}\hat{a}^{\dagger}\hat{a}\ket{\Psi_f} \nonumber \\
&=& \bra{s(r,\varphi)} \hat{S} (r)  \hat{a}^{\dagger}\hat{a} \hat{S}^{\dagger} (r)  \ket{s(r,\varphi)} \nonumber \\
&=&4 \bar{n}(\bar{n}+1) \sin^2{\varphi}. \label{step1}
\eea
In the previous line we have used the fact $\hat{S} (r)  \hat{a}\hat{S}^{\dagger} (r)  = \hat{a}  \cosh{r} + \hat{a}^{\dagger} \sinh{r}$ and $\bar{n}=\sinh^2{r}$. Likewise, we determine the error in this observable to be
\bea
\Delta ^2 (\hat{a}^{\dagger}\hat{a}) &=& \bra{\Psi_f}(\hat{a}^{\dagger}\hat{a})^2\ket{\Psi_f}-\bra{\Psi_f}\hat{a}^{\dagger}\hat{a}\ket{\Psi_f}^2 \nonumber \\
&=& 8 \left[\bar{n} (\bar{n}+1) \sin ^2{\varphi} \right] \times \nonumber \\
&\,& \left[1+2 \bar{n} + 2 \bar{n}^2 - 2\bar{n}(\bar{n}+1) \cos{2\varphi} \right]. \label{step2}
\eea
We find that the measurement signal $\mathcal{S}(\bar{n},\varphi)$ is supersensitive to small fluctuations in $\varphi$:
\bea
\Delta \varphi  = \left[ \frac{\Delta (\hat{a}^{\dagger}\hat{a})}{\big|\frac{\partial}{\partial \varphi} \mathcal{S}(\bar{n},\varphi)\big|} \right]_{\varphi=0}=\frac{1}{ \sqrt{8 \bar{n}+ 8\bar{n}^2}}. \label{step3}
\eea
In fact, this error $\Delta \varphi$ scales better than the $1/\bar{n}$ Heisenberg scaling. In addition, this value of $\Delta \varphi$ saturates the quantum Cram\'{e}r-Rao bound, as $\mathfrak{F}=4\Delta^2 (\hat{a}^{\dagger}\hat{a})=8\bar{n}(\bar{n}+1)$. 

\begin{figure*}
\centering
\includegraphics[width=18 cm,angle=0]{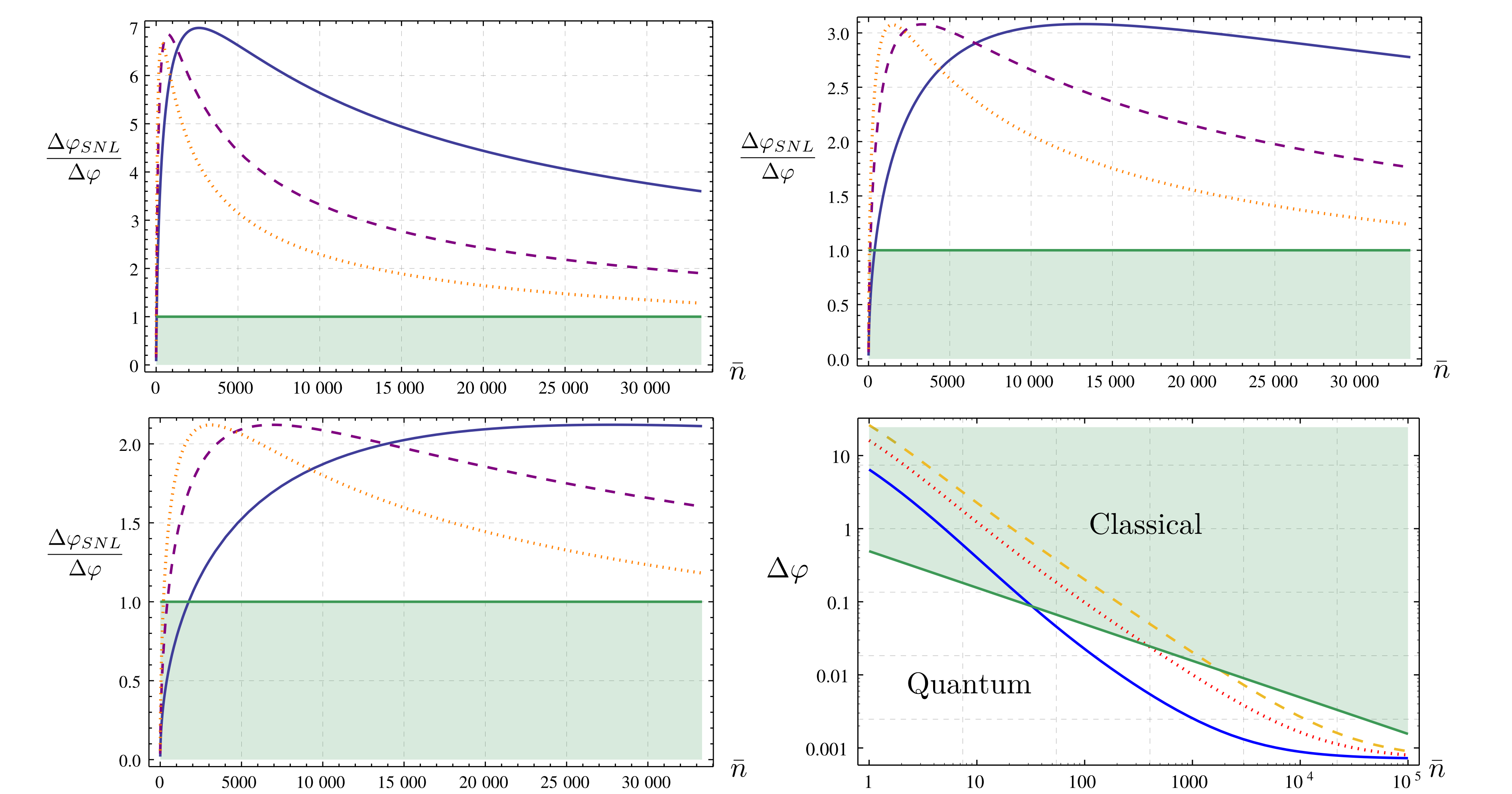}
\caption{(Color online) (a), (b), and (c) depict the ratio of the SNL to the phase error $\Delta \varphi$ of example 2: $\Delta\varphi_{SNL}/\Delta\varphi$, which indicates the extent to which the SNL is beat. This has been plotted for $\varphi=10^{-3}$ (solid, black curve), $\varphi=2 \times 10^{-3}$ (dashed, purple curve), and $\varphi=3 \times 10^{-3}$ (dotted, orange curve).  Figures (a), (b), and (c) correspond to $\eta=0.99$, $\eta=0.95$ and $\eta=0.90$ respectively. The shaded green region corresponds to classical phase sensitivity (below the SNL). Finally, figure (d) shows a log-log plot of $\Delta \varphi$ versus $\bar{n}$ for example 2 when $\varphi = 10^{-3}$. Different values of photon loss are considered: $\eta=0.99$ (solid, blue curve), $\eta=0.95$ (dotted,  red curve), $\eta=0.90$ (dashed, gold curve). The SNL is also depicted (solid, green curve).}
\label{fig}
\end{figure*} 

The setup that we have proposed is most sensitive to losses between $\hat{S}(r)$ and $\hat{S}^{\dagger}(r)$, and after $\hat{S}^{\dagger}(r)$ \cite{PhysRevLett.110.170406}. In order to model the effect of photon loss in this scheme, we insert a fictitious BS with transmissivity $\eta_1$ after the phase shift and another one with transmissivity $\eta_2$ before the intensity detector (see Fig. \ref{fig:scheme1} and Appendix \ref{Losses} for a detailed calculation). Note that we have the freedom to put the $\eta_1$ BS before or after the phase shift because this noise operation commutes with the phase shifter. Additionally, another cause of anti-squeezing inefficiency is mode mismatching. This source of error is also accounted for by the $\eta_1$ BS. After the modified probe state passes through this fictitious BS, we trace over the environment to obtain \cite{PhysRevLett.102.040403} 
\beq
\hat{\sigma}_\varphi =\sum_{j=0}^{\infty} \hat{\Gamma}_j^{(1)} \hat{\rho}_\varphi \hat{\Gamma}_j^{(1) \dagger}, \,\,\, \, \hat{\Gamma}_j^{(1)}=\frac{(1-\eta_1)^{\frac{j}{2}} \eta_{1}^{\frac{1}{2}\hat{a}^{\dagger}\hat{a}}\hat{a}^{j}}{\sqrt{j!}}
\eeq
where $\hat{\rho}_\varphi=\ket{s(r,\varphi)} \bra{s(r,\varphi)}$. Likewise, we obtain the final state, right before the photon detector:
\beq
\hat{\sigma}_f =\sum_{j=0}^{\infty} \hat{\Gamma}_j^{(2)} \hat{\rho}_f \hat{\Gamma}_j^{(2) \dagger}, \,\,\, \, \hat{\Gamma}_j^{(2)}=\frac{(1-\eta_2)^{\frac{j}{2}} \eta_{2}^{\frac{1}{2}\hat{a}^{\dagger}\hat{a}}\hat{a}^{j}}{\sqrt{j!}}
\eeq
with $\hat{\rho}_f= \hat{S}(r) \hat{\sigma}_\varphi \hat{S}^{\dagger}(r)$. In analogy with the lossless case, we can repeat steps (\ref{step1}), (\ref{step2}), and (\ref{step3}); but now the expectation values in these equations should be computed with respect to the density operator $\hat{\sigma}_f$. Due to the Gaussian nature of this protocol, these calculations can be completed simply by considering the second moments of the creation and annihilation operators (see Appendix \ref{Losses}). This yields the noisy signal
\beq
\mathcal{S}(\bar{n},\varphi,\eta)= \bar{n}\eta \left[1+\eta + 2\bar{n}\eta-2(\bar{n}+1)\eta\cos{(2\varphi)} \right]
\eeq
where we have assumed $\eta=\eta_1=\eta_2$ for simplicity. This leads to a estimation error which depends of $\eta$, $\varphi$ and of course $\bar{n}$:
\bea
\Delta ^2 \varphi &=&\big[\eta ^3+2 \eta +12 \eta ^3 \bar{n}^3+16 \eta ^3 \bar{n}^2+8 \eta ^2 \bar{n}^2  \nonumber \\ 
&\,& + \, 4 \eta ^3 \bar{n} (\bar{n}+1)^2 \cos (4 \varphi )+6 \eta ^3 \bar{n} \nonumber \\
&\,&- \, 2 \eta  (\bar{n}+1) \left(\eta + 4 \eta ^2 \bar{n} (2 \bar{n}+1)+4 \eta  \bar{n}+1\right) \cos (2 \varphi )  \nonumber \\
&\,&+ \, 6 \eta ^2 \bar{n}+4 \eta  \bar{n}+1\big]\times \left[\frac{\csc ^2(2 \varphi )}{16 \eta ^3 \bar{n} (\bar{n}+1)^2} \right]. \label{pe}
\eea

In order to demonstrate that (\ref{pe}) predicts sub-SNL phase error, we plot the ratio of $\Delta \varphi_{SNL}=1/\sqrt{4\bar{n}}$ to (\ref{pe}): $\Delta\varphi_{SNL}/\Delta\varphi$ for various values of $\eta$ (See Fig. \ref{fig} a, b, and c). Here we have used the fact that the SNL for the single-mode case is equal to $\Delta \varphi_{SNL}=1/\sqrt{4\bar{n}}$, which follows from the Fisher information for a single-mode probe, $\mathfrak{F}=4\Delta^2 (\hat{a}^{\dagger}\hat{a})$. Note that we do not claim that when losses are present one still attains a better than $1/\sqrt{4\bar{n}}$ scaling in the large $\bar{n}$ limit. Rather, as shown in Fig. 3, a significant improvement over the SNL is attained for experimental parameters (i.e. $\bar{n}$ and $\varphi$) within the range of current technological capabilities.

According to Ref. \cite{Demkowicz2012} and Ref. \cite{knysh2014}, only a constant enhancement over the SNL is possible when losses are introduced. We note here that our results respect the bounds derived in these references. All though only a constant enhancement over the SNL is possible, such an enhancement can still be quite substantial for a give experiment (i.e. for given experimental parameters $\bar{n}$ and $\varphi$) as depicted in Fig. \ref{fig}. It should be possible in practice to keep losses low enough in order to observe such a significant enhancement. The Steinberg group has already demonstrated squeezing then anti-squeezing by retro-reflecting with a high efficiency \cite{PhysRevLett.87.123603, PhysRevLett.88.113601, PhysRevLett.89.037904, PhysRevLett.102.020404}. In addition, as proposed in Ref. \cite{PhysRevLett.110.170406}, a value of $\eta=0.99$ should be attainable if squeezing and anti-squeezing occur in the same solid-state system, where the non-linear parts are divided by a nonactive space layer. For $\eta=0.99$, we predict a 5-fold improvement over the SNL when $\bar{n}=1.5\times10^4$. Note that squeezed states with mean photon number greater than $\bar{n}=3.5 \times10^4$ have been reported in Ref.  \cite{PhysRevLett.100.253601}. If $\eta=0.95$, a 3-fold improvement over the SNL should be observable for $\bar{n}=2\times10^4$ (See Fig. \ref{fig}b). Not only does this scheme yield high sensitivity to small changes in $\varphi$, it also provides high resolution as the probe state employed can be created with current capabilities, where $\bar{n}$ is in the order of many thousands of photons  \cite{PhysRevLett.100.253601}.

Finally note that Ref.\cite{Datta2012} and Ref.\cite{knott} discuss other Heisenberg limited single-mode quantum metrology schemes. These schemes require non-conventional probe preparation techniques as they employ tailored superpositions of Fock states or coherent states.


\section{Conclusion}
In summary, we have shown that the metrological power of quantum light is determined by its inter-mode correlations and intra-mode correlations. This yields an alternative to entanglement for leveraging quantum properties, namely non-classical photon statistics, to surpass classical limitations. 
This is exemplified by our experimental proposal, which employs standard elements of passive and active linear optics, while saturating the quantum Cram\'{e}r-Rao bound with a simple intensity measurement. This proposal performs well in the presence of realistic losses and generates bright probe states (i.e. $\bar{n}$ in the order of many thousands of photons with current technological capabilities), thus overcoming some major obstacles of supersensitive and super-resolving measurement implementation.

\section*{Acknowledgements}

This work was funded by Natural Sciences and Engineering Research Council of Canada. The authors would like to thank D. F. V. James, A. M. Steinberg, and D. H. Mahler for helpful discussions.

\appendix

\section{Particle and mode entanglement}\label{app0}
We illustrated in Sec. (\ref{qfi}) of this manuscript that mode entanglement is not required to beat the SNL and obtain a quantum advantage in metrology. However, there are other types of entanglement that can be studied, including particle entanglement. It can be shown that particle entanglement is necessary but not sufficient for quantum-enhanced interferometry \footnote{In preparation}. 

The twin single-mode squeezed vacuum state discussed at the end of Sec. (\ref{qfi}) does not possess any mode entanglement. Nevertheless as discussed in Ref. \cite{Raf2014} it does possess particle entanglement. Note that particle entanglement needs to be defined in subspaces with a definite number of particles. For instance if the twin single-mode squeezed vacuum state were projected into the two particle subspace and normalized one would obtain
\begin{eqnarray}
\Pi_{2}\ket{s(r,0)}\otimes\ket{s(r,0)} \propto \frac{\ket{2}\otimes \ket{0}+\ket{0}\otimes\ket{2}}{\sqrt{2}},
\end{eqnarray}
were $\Pi_2$ is the projector into the two particle subspace.
Note that the notation used in the previous equation is that of occupation numbers in second quantization, \emph{e.g.}, the last state is a linear combination of a state with particles in arm (mode) 1 and two particles in arm (mode) 2. This state could also be equivalently written in first quantization notation as:
\begin{eqnarray}
\frac{\ket{2}\otimes \ket{0}+\ket{0}\otimes\ket{2}}{\sqrt{2}} = \frac{\ket{a}_1 \otimes \ket{a}_2+\ket{b}_1 \otimes \ket{b}_2}{\sqrt{2}}
\end{eqnarray}
where the notation $\ket{x}_1 \otimes \ket{y}_2$ indicates particle $1$ in mode $x$ and particle 2 in mode $y$. Note that the last state is entangled with respect to the partition $1|2$ and that it is of course symmetrized with respect to the indices 1 and 2.

\section{Losses}\label{Losses}
In this appendix we provide details of the calculation of the effects of losses in the metrology scheme discussed in the main text. We note that the initial state, vacuum, and the operations involved in its transformation, squeezing, amplitude damping (photon loss) and rotations in phase space are gaussian; hence we only need to propagate the first and second moments of the quadratures of the electric field to completely specify the state. Furthermore since there are no displacement operators involved in the scheme the first moments of the state are always zero and we do not need to calculate them.
We remind the reader that a squeezing operator transforms operators according to
\begin{eqnarray}
\hat a_{\text{OUT}}=\hat S(r)^\dagger\hat a_{\text{IN}} \hat S(r)=\hat a_{\text{IN}}\cosh(r)-\hat a_{\text{IN}}^\dagger \sinh(r),
\end{eqnarray}
rotation operators transform the mode according to
\begin{eqnarray}
\hat a_{\text{OUT}}=\hat U(\varphi)^\dagger \hat a_{\text{IN}} \hat U(\varphi)=\hat a_{\text{IN}} e^{-i \varphi}.
\end{eqnarray}
Finally, amplitude damping can be modeled by sending the mode of interest through a beam splitter of amplitude transmissivity $\sqrt{\eta}$ in which the second input port of the beam splitter is in the vacuum state and looking at the output of the transmitted arm \emph{ignoring (tracing out)} the reflected output mode. The effect of a beam splitter on the quadrature operator is
\begin{eqnarray}
\hat a_{\text{OUT}}=\hat U(\eta)^\dagger \hat a_{\text{IN}} \hat U(\eta)=\sqrt{\eta} \hat a_{\text{IN}} +\sqrt{1-\eta} \hat b_{\text{IN}}.
\end{eqnarray}
In the last equation $\hat b_{\text{IN}}$ is the second input mode of the beam splitter that later will be assumed to be prepared in vacuum and traced out.

Because all the operations described before can be implemented using passive and active linear optics the output operators are linear combinations of the input operators. This also implies that the second moments of the outputs can be written as linear combinations of the second moments of the inputs. To use this property let us define the following vector:
\begin{eqnarray}
\mathbf{v}=(\bket{\hat a^2},\bket{(\hat a^\dagger)^2},\bket{\hat a^\dagger \hat a})^T.
\end{eqnarray}
Each of the transformations described before can be represented as an affine (linear transformations augmented with translations) of the vector $\mathbf{v}$: $\mathbf{v}_{\text{OUT}}=\mathbf{M} \mathbf{v}_{\text{IN}}+\mathbf{f}$.
Explicitly, for squeezing we have:
\begin{eqnarray}
\mathbf{A}_{\text{s}}(r)&=&\left(
\begin{array}{ccc}
 c_r^2 & s_r^2 & -s_{2 r} \\
  s_r^2 & c_r^2 & -s_{2 r} \\
 c_r s_r &  c_r s_r & c_{2 r} 
\end{array}
\right),\\
\mathbf{e}(r)&=&\left(-c_r s_r,-c_r s_r,s_r^2 \right),\\
c_r&=&\cosh(r), \quad s_r=\sinh(r),
\end{eqnarray}
for rotations we have
\begin{eqnarray}
\mathbf{B}(\varphi)&=&\text{diag}\left(e^{-2 i \varphi} , e^{2 i \varphi},  1 \right),\\
\mathbf{f}(\varphi)&=&\mathbf{0},
\end{eqnarray}
finally, for amplitude damping we have
\begin{eqnarray}
\mathbf{C}(\eta)&=& \eta\mathbb{I}_3,\\
\mathbf{g}(\eta)&=&\mathbf{0},
\end{eqnarray}
$\text{diag}(x_1, x_2, x_3)$ is a diagonal matrix with entries $x_1,x_2,x_3$ and $\mathbb{I}_3$ is the $3\times 3$ identity matrix.

To obtain the second moments we only need to apply the operations in the right order to the second moments of vacuum, $\mathbf{v}_{\text{IN}}=(0,0,0)$,
\begin{eqnarray}
&&\mathbf{v}_{\text{OUT}}\\
&&=\mathbf{C}(\eta)\left[\mathbf{A}(-r)\left\{\mathbf{C}(\eta) \mathbf{B}(\varphi)\left(\mathbf{A}(r)\mathbf{v}_{\text{IN}}+\mathbf{e}(r) \right)\right\}+\mathbf{e}(-r)\right]\nonumber\\
&&=\mathbf{C}(\eta)\left[\mathbf{A}(-r)\left\{\mathbf{C}(\eta) \mathbf{B}(\varphi)\mathbf{e}(r)\right\}+\mathbf{e}(-r)\right]\nonumber.
\end{eqnarray}
In the last equation $\mathbf{v}_{\text{IN}}$ are the second moments of the state before the protocol is applied, which are all zero since that state is vacuum; likewise $\mathbf{v}_{\text{OUT}}$ are the second moments of the state after the protocol has been applied, explicitly, they are
\begin{eqnarray}
\bket{\hat a^\dagger \hat a}&=&\eta  \bar n [1+\eta+2 \eta \bar n-2(\bar n +1)\eta \cos(2 \varphi)]\\
\bket{a^2}&=&\bket{(a^\dagger)^2}^*=\eta  \sqrt{\bar{n}} \sqrt{\bar{n}+1} \big(\eta  \bar{n} \left(2-e^{2 i \varphi }\right) \nonumber \\ 
&&\quad \quad \quad \quad  -\eta  (\bar{n}+1) e^{-2 i \varphi }+1\big) 
\end{eqnarray}
with $\bar n=\sinh^2 r$. 
Among the second moments we obtain the value of the signal in the metrology protocol, $\bket{\hat a^\dagger \hat a}$. To evaluate the error in the protocol we also calculate the variance of this observable, 
\begin{eqnarray}
\Delta^2 \hat a^\dagger \hat a &=& \bket{(\hat a^\dagger \hat a)^2}-\bket{\hat a^\dagger \hat a}^2\\
&=&\bket{\hat a^\dagger \hat a^\dagger \hat a \hat a}+\bket{\hat a^\dagger \hat a}-\bket{\hat a^\dagger \hat a}^2. \nonumber
\end{eqnarray}
To finalize the calculation we note that for a gaussian state all the normal ordered moments can be expressed as functions of the first and second moments that we already calculated, and that in particular
\begin{eqnarray}
\bket{\hat a^\dagger \hat a^\dagger \hat a \hat a}=2 \bket{\hat a^\dagger \hat a}^2+\bket{\hat a^2} \bket{ \left(\hat a^\dagger\right)^2}.
\end{eqnarray}

\bibliography{Q_metrology_without_entanglement}

\begin{thebibliography}{63}%
\makeatletter
\providecommand \@ifxundefined [1]{%
 \@ifx{#1\undefined}
}%
\providecommand \@ifnum [1]{%
 \ifnum #1\expandafter \@firstoftwo
 \else \expandafter \@secondoftwo
 \fi
}%
\providecommand \@ifx [1]{%
 \ifx #1\expandafter \@firstoftwo
 \else \expandafter \@secondoftwo
 \fi
}%
\providecommand \natexlab [1]{#1}%
\providecommand \enquote  [1]{``#1''}%
\providecommand \bibnamefont  [1]{#1}%
\providecommand \bibfnamefont [1]{#1}%
\providecommand \citenamefont [1]{#1}%
\providecommand \href@noop [0]{\@secondoftwo}%
\providecommand \href [0]{\begingroup \@sanitize@url \@href}%
\providecommand \@href[1]{\@@startlink{#1}\@@href}%
\providecommand \@@href[1]{\endgroup#1\@@endlink}%
\providecommand \@sanitize@url [0]{\catcode `\\12\catcode `\$12\catcode
  `\&12\catcode `\#12\catcode `\^12\catcode `\_12\catcode `\%12\relax}%
\providecommand \@@startlink[1]{}%
\providecommand \@@endlink[0]{}%
\providecommand \url  [0]{\begingroup\@sanitize@url \@url }%
\providecommand \@url [1]{\endgroup\@href {#1}{\urlprefix }}%
\providecommand \urlprefix  [0]{URL }%
\providecommand \Eprint [0]{\href }%
\providecommand \doibase [0]{http://dx.doi.org/}%
\providecommand \selectlanguage [0]{\@gobble}%
\providecommand \bibinfo  [0]{\@secondoftwo}%
\providecommand \bibfield  [0]{\@secondoftwo}%
\providecommand \translation [1]{[#1]}%
\providecommand \BibitemOpen [0]{}%
\providecommand \bibitemStop [0]{}%
\providecommand \bibitemNoStop [0]{.\EOS\space}%
\providecommand \EOS [0]{\spacefactor3000\relax}%
\providecommand \BibitemShut  [1]{\csname bibitem#1\endcsname}%
\let\auto@bib@innerbib\@empty
\bibitem [{\citenamefont {Collaboration}(2011)}]{LIGO11}%
  \BibitemOpen
  \bibfield  {author} {\bibinfo {author} {\bibfnamefont {T.~L.~S.}\
  \bibnamefont {Collaboration}},\ }\href {http://dx.doi.org/10.1038/nphys2083}
  {\bibfield  {journal} {\bibinfo  {journal} {Nat Phys}\ }\textbf {\bibinfo
  {volume} {7}},\ \bibinfo {pages} {962} (\bibinfo {year} {2011})}\BibitemShut
  {NoStop}%
\bibitem [{\citenamefont {Cronin}\ \emph {et~al.}(2009)\citenamefont {Cronin},
  \citenamefont {Schmiedmayer},\ and\ \citenamefont
  {Pritchard}}]{RevModPhys.81.1051}%
  \BibitemOpen
  \bibfield  {author} {\bibinfo {author} {\bibfnamefont {A.~D.}\ \bibnamefont
  {Cronin}}, \bibinfo {author} {\bibfnamefont {J.}~\bibnamefont
  {Schmiedmayer}}, \ and\ \bibinfo {author} {\bibfnamefont {D.~E.}\
  \bibnamefont {Pritchard}},\ }\href {\doibase 10.1103/RevModPhys.81.1051}
  {\bibfield  {journal} {\bibinfo  {journal} {Rev. Mod. Phys.}\ }\textbf
  {\bibinfo {volume} {81}},\ \bibinfo {pages} {1051} (\bibinfo {year}
  {2009})}\BibitemShut {NoStop}%
\bibitem [{\citenamefont {Pavel}\ \emph {et~al.}(2013)\citenamefont {Pavel},
  \citenamefont {Jinga}, \citenamefont {Andronescu}, \citenamefont {Vasile},
  \citenamefont {Kada}, \citenamefont {Sasahara}, \citenamefont {Tosa},
  \citenamefont {Matei}, \citenamefont {Dinescu}, \citenamefont {Dinescu},\
  and\ \citenamefont {Vasile}}]{Pavel2013259}%
  \BibitemOpen
  \bibfield  {author} {\bibinfo {author} {\bibfnamefont {E.}~\bibnamefont
  {Pavel}}, \bibinfo {author} {\bibfnamefont {S.}~\bibnamefont {Jinga}},
  \bibinfo {author} {\bibfnamefont {E.}~\bibnamefont {Andronescu}}, \bibinfo
  {author} {\bibfnamefont {B.}~\bibnamefont {Vasile}}, \bibinfo {author}
  {\bibfnamefont {G.}~\bibnamefont {Kada}}, \bibinfo {author} {\bibfnamefont
  {A.}~\bibnamefont {Sasahara}}, \bibinfo {author} {\bibfnamefont
  {N.}~\bibnamefont {Tosa}}, \bibinfo {author} {\bibfnamefont {A.}~\bibnamefont
  {Matei}}, \bibinfo {author} {\bibfnamefont {M.}~\bibnamefont {Dinescu}},
  \bibinfo {author} {\bibfnamefont {A.}~\bibnamefont {Dinescu}}, \ and\
  \bibinfo {author} {\bibfnamefont {O.}~\bibnamefont {Vasile}},\ }\href
  {\doibase http://dx.doi.org/10.1016/j.optcom.2012.10.079} {\bibfield
  {journal} {\bibinfo  {journal} {Optics Communications}\ }\textbf {\bibinfo
  {volume} {291}},\ \bibinfo {pages} {259 } (\bibinfo {year}
  {2013})}\BibitemShut {NoStop}%
\bibitem [{\citenamefont {Giovannetti}\ \emph {et~al.}(2011)\citenamefont
  {Giovannetti}, \citenamefont {Lloyd},\ and\ \citenamefont
  {Maccone}}]{Giovannetti:2011ys}%
  \BibitemOpen
  \bibfield  {author} {\bibinfo {author} {\bibfnamefont {V.}~\bibnamefont
  {Giovannetti}}, \bibinfo {author} {\bibfnamefont {S.}~\bibnamefont {Lloyd}},
  \ and\ \bibinfo {author} {\bibfnamefont {L.}~\bibnamefont {Maccone}},\ }\href
  {http://dx.doi.org/10.1038/nphoton.2011.35} {\bibfield  {journal} {\bibinfo
  {journal} {Nat Photon}\ }\textbf {\bibinfo {volume} {5}},\ \bibinfo {pages}
  {222} (\bibinfo {year} {2011})}\BibitemShut {NoStop}%
\bibitem [{\citenamefont {Giovannetti}\ \emph {et~al.}(2004)\citenamefont
  {Giovannetti}, \citenamefont {Lloyd},\ and\ \citenamefont
  {Maccone}}]{giovannetti04}%
  \BibitemOpen
  \bibfield  {author} {\bibinfo {author} {\bibfnamefont {V.}~\bibnamefont
  {Giovannetti}}, \bibinfo {author} {\bibfnamefont {S.}~\bibnamefont {Lloyd}},
  \ and\ \bibinfo {author} {\bibfnamefont {L.}~\bibnamefont {Maccone}},\
  }\href@noop {} {\bibfield  {journal} {\bibinfo  {journal} {Science}\ }\textbf
  {\bibinfo {volume} {306}},\ \bibinfo {pages} {1330} (\bibinfo {year}
  {2004})}\BibitemShut {NoStop}%
\bibitem [{Note1()}]{Note1}%
  \BibitemOpen
  \bibinfo {note} {Note that this can be generalized to mixed states, however
  we will mainly be concerned with pure states in this article.}\BibitemShut
  {Stop}%
\bibitem [{\citenamefont {Caves}(1981)}]{Caves81}%
  \BibitemOpen
  \bibfield  {author} {\bibinfo {author} {\bibfnamefont {C.~M.}\ \bibnamefont
  {Caves}},\ }\href {\doibase 10.1103/PhysRevD.23.1693} {\bibfield  {journal}
  {\bibinfo  {journal} {Phys. Rev. D}\ }\textbf {\bibinfo {volume} {23}},\
  \bibinfo {pages} {1693} (\bibinfo {year} {1981})}\BibitemShut {NoStop}%
\bibitem [{\citenamefont {Gerry}(2000)}]{PhysRevA.61.043811}%
  \BibitemOpen
  \bibfield  {author} {\bibinfo {author} {\bibfnamefont {C.~C.}\ \bibnamefont
  {Gerry}},\ }\href {\doibase 10.1103/PhysRevA.61.043811} {\bibfield  {journal}
  {\bibinfo  {journal} {Phys. Rev. A}\ }\textbf {\bibinfo {volume} {61}},\
  \bibinfo {pages} {043811} (\bibinfo {year} {2000})}\BibitemShut {NoStop}%
\bibitem [{\citenamefont {Boto}\ \emph {et~al.}(2000)\citenamefont {Boto},
  \citenamefont {Kok}, \citenamefont {Abrams}, \citenamefont {Braunstein},
  \citenamefont {Williams},\ and\ \citenamefont
  {Dowling}}]{PhysRevLett.85.2733}%
  \BibitemOpen
  \bibfield  {author} {\bibinfo {author} {\bibfnamefont {A.~N.}\ \bibnamefont
  {Boto}}, \bibinfo {author} {\bibfnamefont {P.}~\bibnamefont {Kok}}, \bibinfo
  {author} {\bibfnamefont {D.~S.}\ \bibnamefont {Abrams}}, \bibinfo {author}
  {\bibfnamefont {S.~L.}\ \bibnamefont {Braunstein}}, \bibinfo {author}
  {\bibfnamefont {C.~P.}\ \bibnamefont {Williams}}, \ and\ \bibinfo {author}
  {\bibfnamefont {J.~P.}\ \bibnamefont {Dowling}},\ }\href {\doibase
  10.1103/PhysRevLett.85.2733} {\bibfield  {journal} {\bibinfo  {journal}
  {Phys. Rev. Lett.}\ }\textbf {\bibinfo {volume} {85}},\ \bibinfo {pages}
  {2733} (\bibinfo {year} {2000})}\BibitemShut {NoStop}%
\bibitem [{\citenamefont {Gerry}\ \emph {et~al.}(2002)\citenamefont {Gerry},
  \citenamefont {Benmoussa},\ and\ \citenamefont
  {Campos}}]{PhysRevA.66.013804}%
  \BibitemOpen
  \bibfield  {author} {\bibinfo {author} {\bibfnamefont {C.~C.}\ \bibnamefont
  {Gerry}}, \bibinfo {author} {\bibfnamefont {A.}~\bibnamefont {Benmoussa}}, \
  and\ \bibinfo {author} {\bibfnamefont {R.~A.}\ \bibnamefont {Campos}},\
  }\href {\doibase 10.1103/PhysRevA.66.013804} {\bibfield  {journal} {\bibinfo
  {journal} {Phys. Rev. A}\ }\textbf {\bibinfo {volume} {66}},\ \bibinfo
  {pages} {013804} (\bibinfo {year} {2002})}\BibitemShut {NoStop}%
\bibitem [{\citenamefont {Lee}\ \emph {et~al.}(2002)\citenamefont {Lee},
  \citenamefont {Kok},\ and\ \citenamefont
  {Dowling}}]{doi:10.1080/0950034021000011536}%
  \BibitemOpen
  \bibfield  {author} {\bibinfo {author} {\bibfnamefont {H.}~\bibnamefont
  {Lee}}, \bibinfo {author} {\bibfnamefont {P.}~\bibnamefont {Kok}}, \ and\
  \bibinfo {author} {\bibfnamefont {J.~P.}\ \bibnamefont {Dowling}},\ }\href
  {\doibase 10.1080/0950034021000011536} {\bibfield  {journal} {\bibinfo
  {journal} {Journal of Modern Optics}\ }\textbf {\bibinfo {volume} {49}},\
  \bibinfo {pages} {2325} (\bibinfo {year} {2002})}\BibitemShut {NoStop}%
\bibitem [{\citenamefont {Sanders}\ and\ \citenamefont
  {Milburn}(1995)}]{PhysRevLett.75.2944}%
  \BibitemOpen
  \bibfield  {author} {\bibinfo {author} {\bibfnamefont {B.~C.}\ \bibnamefont
  {Sanders}}\ and\ \bibinfo {author} {\bibfnamefont {G.~J.}\ \bibnamefont
  {Milburn}},\ }\href {\doibase 10.1103/PhysRevLett.75.2944} {\bibfield
  {journal} {\bibinfo  {journal} {Phys. Rev. Lett.}\ }\textbf {\bibinfo
  {volume} {75}},\ \bibinfo {pages} {2944} (\bibinfo {year}
  {1995})}\BibitemShut {NoStop}%
\bibitem [{\citenamefont {Ou}(1996)}]{PhysRevLett.77.2352}%
  \BibitemOpen
  \bibfield  {author} {\bibinfo {author} {\bibfnamefont {Z.~Y.}\ \bibnamefont
  {Ou}},\ }\href {\doibase 10.1103/PhysRevLett.77.2352} {\bibfield  {journal}
  {\bibinfo  {journal} {Phys. Rev. Lett.}\ }\textbf {\bibinfo {volume} {77}},\
  \bibinfo {pages} {2352} (\bibinfo {year} {1996})}\BibitemShut {NoStop}%
\bibitem [{\citenamefont {Giovannetti}\ \emph {et~al.}(2006)\citenamefont
  {Giovannetti}, \citenamefont {Lloyd},\ and\ \citenamefont
  {Maccone}}]{PhysRevLett.96.010401}%
  \BibitemOpen
  \bibfield  {author} {\bibinfo {author} {\bibfnamefont {V.}~\bibnamefont
  {Giovannetti}}, \bibinfo {author} {\bibfnamefont {S.}~\bibnamefont {Lloyd}},
  \ and\ \bibinfo {author} {\bibfnamefont {L.}~\bibnamefont {Maccone}},\ }\href
  {\doibase 10.1103/PhysRevLett.96.010401} {\bibfield  {journal} {\bibinfo
  {journal} {Phys. Rev. Lett.}\ }\textbf {\bibinfo {volume} {96}},\ \bibinfo
  {pages} {010401} (\bibinfo {year} {2006})}\BibitemShut {NoStop}%
\bibitem [{\citenamefont {Zwierz}\ \emph {et~al.}(2010)\citenamefont {Zwierz},
  \citenamefont {P\'erez-Delgado},\ and\ \citenamefont
  {Kok}}]{PhysRevLett.105.180402}%
  \BibitemOpen
  \bibfield  {author} {\bibinfo {author} {\bibfnamefont {M.}~\bibnamefont
  {Zwierz}}, \bibinfo {author} {\bibfnamefont {C.~A.}\ \bibnamefont
  {P\'erez-Delgado}}, \ and\ \bibinfo {author} {\bibfnamefont {P.}~\bibnamefont
  {Kok}},\ }\href {\doibase 10.1103/PhysRevLett.105.180402} {\bibfield
  {journal} {\bibinfo  {journal} {Phys. Rev. Lett.}\ }\textbf {\bibinfo
  {volume} {105}},\ \bibinfo {pages} {180402} (\bibinfo {year}
  {2010})}\BibitemShut {NoStop}%
\bibitem [{\citenamefont {Hall}\ \emph {et~al.}(2012)\citenamefont {Hall},
  \citenamefont {Berry}, \citenamefont {Zwierz},\ and\ \citenamefont
  {Wiseman}}]{PhysRevA.85.041802}%
  \BibitemOpen
  \bibfield  {author} {\bibinfo {author} {\bibfnamefont {M.~J.~W.}\
  \bibnamefont {Hall}}, \bibinfo {author} {\bibfnamefont {D.~W.}\ \bibnamefont
  {Berry}}, \bibinfo {author} {\bibfnamefont {M.}~\bibnamefont {Zwierz}}, \
  and\ \bibinfo {author} {\bibfnamefont {H.~M.}\ \bibnamefont {Wiseman}},\
  }\href {\doibase 10.1103/PhysRevA.85.041802} {\bibfield  {journal} {\bibinfo
  {journal} {Phys. Rev. A}\ }\textbf {\bibinfo {volume} {85}},\ \bibinfo
  {pages} {041802} (\bibinfo {year} {2012})}\BibitemShut {NoStop}%
\bibitem [{\citenamefont {Giovannetti}\ and\ \citenamefont
  {Maccone}(2012)}]{PhysRevLett.108.210404}%
  \BibitemOpen
  \bibfield  {author} {\bibinfo {author} {\bibfnamefont {V.}~\bibnamefont
  {Giovannetti}}\ and\ \bibinfo {author} {\bibfnamefont {L.}~\bibnamefont
  {Maccone}},\ }\href {\doibase 10.1103/PhysRevLett.108.210404} {\bibfield
  {journal} {\bibinfo  {journal} {Phys. Rev. Lett.}\ }\textbf {\bibinfo
  {volume} {108}},\ \bibinfo {pages} {210404} (\bibinfo {year}
  {2012})}\BibitemShut {NoStop}%
\bibitem [{\citenamefont {Yurke}(1986)}]{PhysRevLett.56.1515}%
  \BibitemOpen
  \bibfield  {author} {\bibinfo {author} {\bibfnamefont {B.}~\bibnamefont
  {Yurke}},\ }\href {\doibase 10.1103/PhysRevLett.56.1515} {\bibfield
  {journal} {\bibinfo  {journal} {Phys. Rev. Lett.}\ }\textbf {\bibinfo
  {volume} {56}},\ \bibinfo {pages} {1515} (\bibinfo {year}
  {1986})}\BibitemShut {NoStop}%
\bibitem [{\citenamefont {Yurke}\ \emph {et~al.}(1986)\citenamefont {Yurke},
  \citenamefont {McCall},\ and\ \citenamefont {Klauder}}]{PhysRevA.33.4033}%
  \BibitemOpen
  \bibfield  {author} {\bibinfo {author} {\bibfnamefont {B.}~\bibnamefont
  {Yurke}}, \bibinfo {author} {\bibfnamefont {S.~L.}\ \bibnamefont {McCall}}, \
  and\ \bibinfo {author} {\bibfnamefont {J.~R.}\ \bibnamefont {Klauder}},\
  }\href {\doibase 10.1103/PhysRevA.33.4033} {\bibfield  {journal} {\bibinfo
  {journal} {Phys. Rev. A}\ }\textbf {\bibinfo {volume} {33}},\ \bibinfo
  {pages} {4033} (\bibinfo {year} {1986})}\BibitemShut {NoStop}%
\bibitem [{\citenamefont {Holland}\ and\ \citenamefont
  {Burnett}(1993)}]{PhysRevLett.71.1355}%
  \BibitemOpen
  \bibfield  {author} {\bibinfo {author} {\bibfnamefont {M.~J.}\ \bibnamefont
  {Holland}}\ and\ \bibinfo {author} {\bibfnamefont {K.}~\bibnamefont
  {Burnett}},\ }\href {\doibase 10.1103/PhysRevLett.71.1355} {\bibfield
  {journal} {\bibinfo  {journal} {Phys. Rev. Lett.}\ }\textbf {\bibinfo
  {volume} {71}},\ \bibinfo {pages} {1355} (\bibinfo {year}
  {1993})}\BibitemShut {NoStop}%
\bibitem [{\citenamefont {Jacobson}\ \emph {et~al.}(1995)\citenamefont
  {Jacobson}, \citenamefont {Bj{\"o}rk},\ and\ \citenamefont
  {Yamamoto}}]{Jacobson:1995fk}%
  \BibitemOpen
  \bibfield  {author} {\bibinfo {author} {\bibfnamefont {J.}~\bibnamefont
  {Jacobson}}, \bibinfo {author} {\bibfnamefont {G.}~\bibnamefont {Bj{\"o}rk}},
  \ and\ \bibinfo {author} {\bibfnamefont {Y.}~\bibnamefont {Yamamoto}},\
  }\href {\doibase 10.1007/BF01135861} {\bibfield  {journal} {\bibinfo
  {journal} {Applied Physics B}\ }\textbf {\bibinfo {volume} {60}},\ \bibinfo
  {pages} {187} (\bibinfo {year} {1995})}\BibitemShut {NoStop}%
\bibitem [{\citenamefont {Bouyer}\ and\ \citenamefont
  {Kasevich}(1997)}]{PhysRevA.56.R1083}%
  \BibitemOpen
  \bibfield  {author} {\bibinfo {author} {\bibfnamefont {P.}~\bibnamefont
  {Bouyer}}\ and\ \bibinfo {author} {\bibfnamefont {M.~A.}\ \bibnamefont
  {Kasevich}},\ }\href {\doibase 10.1103/PhysRevA.56.R1083} {\bibfield
  {journal} {\bibinfo  {journal} {Phys. Rev. A}\ }\textbf {\bibinfo {volume}
  {56}},\ \bibinfo {pages} {R1083} (\bibinfo {year} {1997})}\BibitemShut
  {NoStop}%
\bibitem [{\citenamefont {Pezz\'e}\ and\ \citenamefont
  {Smerzi}(2008)}]{PhysRevLett.100.073601}%
  \BibitemOpen
  \bibfield  {author} {\bibinfo {author} {\bibfnamefont {L.}~\bibnamefont
  {Pezz\'e}}\ and\ \bibinfo {author} {\bibfnamefont {A.}~\bibnamefont
  {Smerzi}},\ }\href {\doibase 10.1103/PhysRevLett.100.073601} {\bibfield
  {journal} {\bibinfo  {journal} {Phys. Rev. Lett.}\ }\textbf {\bibinfo
  {volume} {100}},\ \bibinfo {pages} {073601} (\bibinfo {year}
  {2008})}\BibitemShut {NoStop}%
\bibitem [{\citenamefont {Pezz\'e}\ and\ \citenamefont
  {Smerzi}(2013)}]{PhysRevLett.110.163604}%
  \BibitemOpen
  \bibfield  {author} {\bibinfo {author} {\bibfnamefont {L.}~\bibnamefont
  {Pezz\'e}}\ and\ \bibinfo {author} {\bibfnamefont {A.}~\bibnamefont
  {Smerzi}},\ }\href {\doibase 10.1103/PhysRevLett.110.163604} {\bibfield
  {journal} {\bibinfo  {journal} {Phys. Rev. Lett.}\ }\textbf {\bibinfo
  {volume} {110}},\ \bibinfo {pages} {163604} (\bibinfo {year}
  {2013})}\BibitemShut {NoStop}%
\bibitem [{\citenamefont {Gilbert}\ \emph {et~al.}(2008)\citenamefont
  {Gilbert}, \citenamefont {Hamrick},\ and\ \citenamefont
  {Weinstein}}]{Gilbert:08}%
  \BibitemOpen
  \bibfield  {author} {\bibinfo {author} {\bibfnamefont {G.}~\bibnamefont
  {Gilbert}}, \bibinfo {author} {\bibfnamefont {M.}~\bibnamefont {Hamrick}}, \
  and\ \bibinfo {author} {\bibfnamefont {Y.~S.}\ \bibnamefont {Weinstein}},\
  }\href {\doibase 10.1364/JOSAB.25.001336} {\bibfield  {journal} {\bibinfo
  {journal} {J. Opt. Soc. Am. B}\ }\textbf {\bibinfo {volume} {25}},\ \bibinfo
  {pages} {1336} (\bibinfo {year} {2008})}\BibitemShut {NoStop}%
\bibitem [{\citenamefont {Lee}\ \emph {et~al.}(2009)\citenamefont {Lee},
  \citenamefont {Huver}, \citenamefont {Lee}, \citenamefont {Kaplan},
  \citenamefont {McCracken}, \citenamefont {Min}, \citenamefont {Uskov},
  \citenamefont {Wildfeuer}, \citenamefont {Veronis},\ and\ \citenamefont
  {Dowling}}]{PhysRevA.80.063803}%
  \BibitemOpen
  \bibfield  {author} {\bibinfo {author} {\bibfnamefont {T.-W.}\ \bibnamefont
  {Lee}}, \bibinfo {author} {\bibfnamefont {S.~D.}\ \bibnamefont {Huver}},
  \bibinfo {author} {\bibfnamefont {H.}~\bibnamefont {Lee}}, \bibinfo {author}
  {\bibfnamefont {L.}~\bibnamefont {Kaplan}}, \bibinfo {author} {\bibfnamefont
  {S.~B.}\ \bibnamefont {McCracken}}, \bibinfo {author} {\bibfnamefont
  {C.}~\bibnamefont {Min}}, \bibinfo {author} {\bibfnamefont {D.~B.}\
  \bibnamefont {Uskov}}, \bibinfo {author} {\bibfnamefont {C.~F.}\ \bibnamefont
  {Wildfeuer}}, \bibinfo {author} {\bibfnamefont {G.}~\bibnamefont {Veronis}},
  \ and\ \bibinfo {author} {\bibfnamefont {J.~P.}\ \bibnamefont {Dowling}},\
  }\href {\doibase 10.1103/PhysRevA.80.063803} {\bibfield  {journal} {\bibinfo
  {journal} {Phys. Rev. A}\ }\textbf {\bibinfo {volume} {80}},\ \bibinfo
  {pages} {063803} (\bibinfo {year} {2009})}\BibitemShut {NoStop}%
\bibitem [{\citenamefont {Dorner}\ \emph {et~al.}(2009)\citenamefont {Dorner},
  \citenamefont {Demkowicz-Dobrzanski}, \citenamefont {Smith}, \citenamefont
  {Lundeen}, \citenamefont {Wasilewski}, \citenamefont {Banaszek},\ and\
  \citenamefont {Walmsley}}]{PhysRevLett.102.040403}%
  \BibitemOpen
  \bibfield  {author} {\bibinfo {author} {\bibfnamefont {U.}~\bibnamefont
  {Dorner}}, \bibinfo {author} {\bibfnamefont {R.}~\bibnamefont
  {Demkowicz-Dobrzanski}}, \bibinfo {author} {\bibfnamefont {B.~J.}\
  \bibnamefont {Smith}}, \bibinfo {author} {\bibfnamefont {J.~S.}\ \bibnamefont
  {Lundeen}}, \bibinfo {author} {\bibfnamefont {W.}~\bibnamefont {Wasilewski}},
  \bibinfo {author} {\bibfnamefont {K.}~\bibnamefont {Banaszek}}, \ and\
  \bibinfo {author} {\bibfnamefont {I.~A.}\ \bibnamefont {Walmsley}},\ }\href
  {\doibase 10.1103/PhysRevLett.102.040403} {\bibfield  {journal} {\bibinfo
  {journal} {Phys. Rev. Lett.}\ }\textbf {\bibinfo {volume} {102}},\ \bibinfo
  {pages} {040403} (\bibinfo {year} {2009})}\BibitemShut {NoStop}%
\bibitem [{\citenamefont {Knysh}\ \emph {et~al.}(2011)\citenamefont {Knysh},
  \citenamefont {Smelyanskiy},\ and\ \citenamefont
  {Durkin}}]{PhysRevA.83.021804}%
  \BibitemOpen
  \bibfield  {author} {\bibinfo {author} {\bibfnamefont {S.}~\bibnamefont
  {Knysh}}, \bibinfo {author} {\bibfnamefont {V.~N.}\ \bibnamefont
  {Smelyanskiy}}, \ and\ \bibinfo {author} {\bibfnamefont {G.~A.}\ \bibnamefont
  {Durkin}},\ }\href {\doibase 10.1103/PhysRevA.83.021804} {\bibfield
  {journal} {\bibinfo  {journal} {Phys. Rev. A}\ }\textbf {\bibinfo {volume}
  {83}},\ \bibinfo {pages} {021804} (\bibinfo {year} {2011})}\BibitemShut
  {NoStop}%
\bibitem [{\citenamefont {Escher}\ \emph
  {et~al.}(2011{\natexlab{a}})\citenamefont {Escher}, \citenamefont
  {de~Matos~Filho},\ and\ \citenamefont {Davidovich}}]{Escher:2011zr}%
  \BibitemOpen
  \bibfield  {author} {\bibinfo {author} {\bibfnamefont {B.~M.}\ \bibnamefont
  {Escher}}, \bibinfo {author} {\bibfnamefont {R.~L.}\ \bibnamefont
  {de~Matos~Filho}}, \ and\ \bibinfo {author} {\bibfnamefont {L.}~\bibnamefont
  {Davidovich}},\ }\href {http://dx.doi.org/10.1038/nphys1958} {\bibfield
  {journal} {\bibinfo  {journal} {Nat Phys}\ }\textbf {\bibinfo {volume} {7}},\
  \bibinfo {pages} {406} (\bibinfo {year} {2011}{\natexlab{a}})}\BibitemShut
  {NoStop}%
\bibitem [{\citenamefont {Escher}\ \emph
  {et~al.}(2011{\natexlab{b}})\citenamefont {Escher}, \citenamefont
  {de~Matos~Filho},\ and\ \citenamefont {Davidovich}}]{escher2011quantum}%
  \BibitemOpen
  \bibfield  {author} {\bibinfo {author} {\bibfnamefont {B.}~\bibnamefont
  {Escher}}, \bibinfo {author} {\bibfnamefont {R.}~\bibnamefont
  {de~Matos~Filho}}, \ and\ \bibinfo {author} {\bibfnamefont {L.}~\bibnamefont
  {Davidovich}},\ }\href@noop {} {\bibfield  {journal} {\bibinfo  {journal}
  {Brazilian Journal of Physics}\ }\textbf {\bibinfo {volume} {41}},\ \bibinfo
  {pages} {229} (\bibinfo {year} {2011}{\natexlab{b}})}\BibitemShut {NoStop}%
\bibitem [{\citenamefont {Huver}\ \emph {et~al.}(2008)\citenamefont {Huver},
  \citenamefont {Wildfeuer},\ and\ \citenamefont
  {Dowling}}]{PhysRevA.78.063828}%
  \BibitemOpen
  \bibfield  {author} {\bibinfo {author} {\bibfnamefont {S.~D.}\ \bibnamefont
  {Huver}}, \bibinfo {author} {\bibfnamefont {C.~F.}\ \bibnamefont
  {Wildfeuer}}, \ and\ \bibinfo {author} {\bibfnamefont {J.~P.}\ \bibnamefont
  {Dowling}},\ }\href {\doibase 10.1103/PhysRevA.78.063828} {\bibfield
  {journal} {\bibinfo  {journal} {Phys. Rev. A}\ }\textbf {\bibinfo {volume}
  {78}},\ \bibinfo {pages} {063828} (\bibinfo {year} {2008})}\BibitemShut
  {NoStop}%
\bibitem [{\citenamefont {Sahota}\ and\ \citenamefont
  {James}(2013)}]{sahota13}%
  \BibitemOpen
  \bibfield  {author} {\bibinfo {author} {\bibfnamefont {J.}~\bibnamefont
  {Sahota}}\ and\ \bibinfo {author} {\bibfnamefont {D.~F.~V.}\ \bibnamefont
  {James}},\ }\href {\doibase 10.1103/PhysRevA.88.063820} {\bibfield  {journal}
  {\bibinfo  {journal} {Phys. Rev. A}\ }\textbf {\bibinfo {volume} {88}},\
  \bibinfo {pages} {063820} (\bibinfo {year} {2013})}\BibitemShut {NoStop}%
\bibitem [{\citenamefont {Nielsen}\ and\ \citenamefont
  {Chuang}(2010)}]{nielsen2010quantum}%
  \BibitemOpen
  \bibfield  {author} {\bibinfo {author} {\bibfnamefont {M.~A.}\ \bibnamefont
  {Nielsen}}\ and\ \bibinfo {author} {\bibfnamefont {I.~L.}\ \bibnamefont
  {Chuang}},\ }\href@noop {} {\emph {\bibinfo {title} {Quantum computation and
  quantum information}}}\ (\bibinfo  {publisher} {Cambridge university press},\
  \bibinfo {year} {2010})\BibitemShut {NoStop}%
\bibitem [{\citenamefont {Bennett}\ \emph {et~al.}(1993)\citenamefont
  {Bennett}, \citenamefont {Brassard}, \citenamefont {Cr\'epeau}, \citenamefont
  {Jozsa}, \citenamefont {Peres},\ and\ \citenamefont
  {Wootters}}]{PhysRevLett.70.1895}%
  \BibitemOpen
  \bibfield  {author} {\bibinfo {author} {\bibfnamefont {C.~H.}\ \bibnamefont
  {Bennett}}, \bibinfo {author} {\bibfnamefont {G.}~\bibnamefont {Brassard}},
  \bibinfo {author} {\bibfnamefont {C.}~\bibnamefont {Cr\'epeau}}, \bibinfo
  {author} {\bibfnamefont {R.}~\bibnamefont {Jozsa}}, \bibinfo {author}
  {\bibfnamefont {A.}~\bibnamefont {Peres}}, \ and\ \bibinfo {author}
  {\bibfnamefont {W.~K.}\ \bibnamefont {Wootters}},\ }\href {\doibase
  10.1103/PhysRevLett.70.1895} {\bibfield  {journal} {\bibinfo  {journal}
  {Phys. Rev. Lett.}\ }\textbf {\bibinfo {volume} {70}},\ \bibinfo {pages}
  {1895} (\bibinfo {year} {1993})}\BibitemShut {NoStop}%
\bibitem [{\citenamefont {Gisin}\ \emph {et~al.}(2002)\citenamefont {Gisin},
  \citenamefont {Ribordy}, \citenamefont {Tittel},\ and\ \citenamefont
  {Zbinden}}]{RevModPhys.74.145}%
  \BibitemOpen
  \bibfield  {author} {\bibinfo {author} {\bibfnamefont {N.}~\bibnamefont
  {Gisin}}, \bibinfo {author} {\bibfnamefont {G.}~\bibnamefont {Ribordy}},
  \bibinfo {author} {\bibfnamefont {W.}~\bibnamefont {Tittel}}, \ and\ \bibinfo
  {author} {\bibfnamefont {H.}~\bibnamefont {Zbinden}},\ }\href {\doibase
  10.1103/RevModPhys.74.145} {\bibfield  {journal} {\bibinfo  {journal} {Rev.
  Mod. Phys.}\ }\textbf {\bibinfo {volume} {74}},\ \bibinfo {pages} {145}
  (\bibinfo {year} {2002})}\BibitemShut {NoStop}%
\bibitem [{\citenamefont {Migdall}\ \emph {et~al.}(2002)\citenamefont
  {Migdall}, \citenamefont {Castelletto}, \citenamefont {Degiovanni},\ and\
  \citenamefont {Rastello}}]{Migdall:02}%
  \BibitemOpen
  \bibfield  {author} {\bibinfo {author} {\bibfnamefont {A.}~\bibnamefont
  {Migdall}}, \bibinfo {author} {\bibfnamefont {S.}~\bibnamefont
  {Castelletto}}, \bibinfo {author} {\bibfnamefont {I.~P.}\ \bibnamefont
  {Degiovanni}}, \ and\ \bibinfo {author} {\bibfnamefont {M.~L.}\ \bibnamefont
  {Rastello}},\ }\href {\doibase 10.1364/AO.41.002914} {\bibfield  {journal}
  {\bibinfo  {journal} {Appl. Opt.}\ }\textbf {\bibinfo {volume} {41}},\
  \bibinfo {pages} {2914} (\bibinfo {year} {2002})}\BibitemShut {NoStop}%
\bibitem [{\citenamefont {Dowling}\ \emph {et~al.}(2006)\citenamefont
  {Dowling}, \citenamefont {Gatti},\ and\ \citenamefont
  {Sergienko}}]{doi:10.1080/09500340600589382}%
  \BibitemOpen
  \bibfield  {author} {\bibinfo {author} {\bibfnamefont {J.~P.}\ \bibnamefont
  {Dowling}}, \bibinfo {author} {\bibfnamefont {A.}~\bibnamefont {Gatti}}, \
  and\ \bibinfo {author} {\bibfnamefont {A.}~\bibnamefont {Sergienko}},\ }\href
  {\doibase 10.1080/09500340600589382} {\bibfield  {journal} {\bibinfo
  {journal} {Journal of Modern Optics}\ }\textbf {\bibinfo {volume} {53}},\
  \bibinfo {pages} {573} (\bibinfo {year} {2006})}\BibitemShut {NoStop}%
\bibitem [{\citenamefont {Lloyd}(2008)}]{Lloyd12092008}%
  \BibitemOpen
  \bibfield  {author} {\bibinfo {author} {\bibfnamefont {S.}~\bibnamefont
  {Lloyd}},\ }\href {\doibase 10.1126/science.1160627} {\bibfield  {journal}
  {\bibinfo  {journal} {Science}\ }\textbf {\bibinfo {volume} {321}},\ \bibinfo
  {pages} {1463} (\bibinfo {year} {2008})}\BibitemShut {NoStop}%
\bibitem [{\citenamefont {Tilma}\ \emph {et~al.}(2010)\citenamefont {Tilma},
  \citenamefont {Hamaji}, \citenamefont {Munro},\ and\ \citenamefont
  {Nemoto}}]{PhysRevA.81.022108}%
  \BibitemOpen
  \bibfield  {author} {\bibinfo {author} {\bibfnamefont {T.}~\bibnamefont
  {Tilma}}, \bibinfo {author} {\bibfnamefont {S.}~\bibnamefont {Hamaji}},
  \bibinfo {author} {\bibfnamefont {W.~J.}\ \bibnamefont {Munro}}, \ and\
  \bibinfo {author} {\bibfnamefont {K.}~\bibnamefont {Nemoto}},\ }\href
  {\doibase 10.1103/PhysRevA.81.022108} {\bibfield  {journal} {\bibinfo
  {journal} {Phys. Rev. A}\ }\textbf {\bibinfo {volume} {81}},\ \bibinfo
  {pages} {022108} (\bibinfo {year} {2010})}\BibitemShut {NoStop}%
\bibitem [{\citenamefont {Datta}\ and\ \citenamefont
  {Shaji}(2012)}]{Datta2012}%
  \BibitemOpen
  \bibfield  {author} {\bibinfo {author} {\bibfnamefont {A.}~\bibnamefont
  {Datta}}\ and\ \bibinfo {author} {\bibfnamefont {A.}~\bibnamefont {Shaji}},\
  }\href@noop {} {\bibfield  {journal} {\bibinfo  {journal} {Mod. Phys. Lett.
  B}\ }\textbf {\bibinfo {volume} {26}} (\bibinfo {year} {2012})}\BibitemShut
  {NoStop}%
\bibitem [{\citenamefont {Joo}\ \emph {et~al.}(2011)\citenamefont {Joo},
  \citenamefont {Munro},\ and\ \citenamefont
  {Spiller}}]{PhysRevLett.107.083601}%
  \BibitemOpen
  \bibfield  {author} {\bibinfo {author} {\bibfnamefont {J.}~\bibnamefont
  {Joo}}, \bibinfo {author} {\bibfnamefont {W.~J.}\ \bibnamefont {Munro}}, \
  and\ \bibinfo {author} {\bibfnamefont {T.~P.}\ \bibnamefont {Spiller}},\
  }\href@noop {} {\bibfield  {journal} {\bibinfo  {journal} {Phys. Rev. Lett.}\
  }\textbf {\bibinfo {volume} {107}},\ \bibinfo {pages} {083601} (\bibinfo
  {year} {2011})}\BibitemShut {NoStop}%
\bibitem [{\citenamefont {Helstrom}(1976)}]{1976quantum}%
  \BibitemOpen
  \bibfield  {author} {\bibinfo {author} {\bibfnamefont {C.~W.}\ \bibnamefont
  {Helstrom}},\ }\href {http://books.google.ca/books?id=Ne3iT\_QLcsMC} {\emph
  {\bibinfo {title} {Quantum detection and estimation theory}}},\ Mathematics
  in Science and Engineering\ (\bibinfo  {publisher} {Elsevier Science},\
  \bibinfo {year} {1976})\BibitemShut {NoStop}%
\bibitem [{\citenamefont {Braunstein}\ and\ \citenamefont
  {Caves}(1994)}]{Caves94}%
  \BibitemOpen
  \bibfield  {author} {\bibinfo {author} {\bibfnamefont {S.~L.}\ \bibnamefont
  {Braunstein}}\ and\ \bibinfo {author} {\bibfnamefont {C.~M.}\ \bibnamefont
  {Caves}},\ }\href {\doibase 10.1103/PhysRevLett.72.3439} {\bibfield
  {journal} {\bibinfo  {journal} {Phys. Rev. Lett.}\ }\textbf {\bibinfo
  {volume} {72}},\ \bibinfo {pages} {3439} (\bibinfo {year}
  {1994})}\BibitemShut {NoStop}%
\bibitem [{\citenamefont {Braunstein}\ \emph {et~al.}(1996)\citenamefont
  {Braunstein}, \citenamefont {Caves},\ and\ \citenamefont
  {Milburn}}]{braunstein1996generalized}%
  \BibitemOpen
  \bibfield  {author} {\bibinfo {author} {\bibfnamefont {S.~L.}\ \bibnamefont
  {Braunstein}}, \bibinfo {author} {\bibfnamefont {C.~M.}\ \bibnamefont
  {Caves}}, \ and\ \bibinfo {author} {\bibfnamefont {G.~J.}\ \bibnamefont
  {Milburn}},\ }\href@noop {} {\bibfield  {journal} {\bibinfo  {journal}
  {Annals of Physics}\ }\textbf {\bibinfo {volume} {247}},\ \bibinfo {pages}
  {135} (\bibinfo {year} {1996})}\BibitemShut {NoStop}%
\bibitem [{\citenamefont {Gerry}\ and\ \citenamefont
  {Knight}(2005)}]{gerry2005introductory}%
  \BibitemOpen
  \bibfield  {author} {\bibinfo {author} {\bibfnamefont {C.}~\bibnamefont
  {Gerry}}\ and\ \bibinfo {author} {\bibfnamefont {P.}~\bibnamefont {Knight}},\
  }\href {http://books.google.ca/books?id=CgByyoBJJwgC} {\emph {\bibinfo
  {title} {Introductory Quantum Optics}}}\ (\bibinfo  {publisher} {Cambridge
  University Press},\ \bibinfo {year} {2005})\BibitemShut {NoStop}%
\bibitem [{\citenamefont {Hofmann}(2009)}]{PhysRevA.79.033822}%
  \BibitemOpen
  \bibfield  {author} {\bibinfo {author} {\bibfnamefont {H.~F.}\ \bibnamefont
  {Hofmann}},\ }\href {\doibase 10.1103/PhysRevA.79.033822} {\bibfield
  {journal} {\bibinfo  {journal} {Phys. Rev. A}\ }\textbf {\bibinfo {volume}
  {79}},\ \bibinfo {pages} {033822} (\bibinfo {year} {2009})}\BibitemShut
  {NoStop}%
\bibitem [{\citenamefont {Anisimov}\ \emph {et~al.}(2010)\citenamefont
  {Anisimov}, \citenamefont {Raterman}, \citenamefont {Chiruvelli},
  \citenamefont {Plick}, \citenamefont {Huver}, \citenamefont {Lee},\ and\
  \citenamefont {Dowling}}]{PhysRevLett.104.103602}%
  \BibitemOpen
  \bibfield  {author} {\bibinfo {author} {\bibfnamefont {P.~M.}\ \bibnamefont
  {Anisimov}}, \bibinfo {author} {\bibfnamefont {G.~M.}\ \bibnamefont
  {Raterman}}, \bibinfo {author} {\bibfnamefont {A.}~\bibnamefont
  {Chiruvelli}}, \bibinfo {author} {\bibfnamefont {W.~N.}\ \bibnamefont
  {Plick}}, \bibinfo {author} {\bibfnamefont {S.~D.}\ \bibnamefont {Huver}},
  \bibinfo {author} {\bibfnamefont {H.}~\bibnamefont {Lee}}, \ and\ \bibinfo
  {author} {\bibfnamefont {J.~P.}\ \bibnamefont {Dowling}},\ }\href {\doibase
  10.1103/PhysRevLett.104.103602} {\bibfield  {journal} {\bibinfo  {journal}
  {Phys. Rev. Lett.}\ }\textbf {\bibinfo {volume} {104}},\ \bibinfo {pages}
  {103602} (\bibinfo {year} {2010})}\BibitemShut {NoStop}%
\bibitem [{\citenamefont {Lang}\ and\ \citenamefont {Caves}(2014)}]{Lang2014}%
  \BibitemOpen
  \bibfield  {author} {\bibinfo {author} {\bibfnamefont {M.~D.}\ \bibnamefont
  {Lang}}\ and\ \bibinfo {author} {\bibfnamefont {C.~M.}\ \bibnamefont
  {Caves}},\ }\href@noop {} {\bibfield  {journal} {\bibinfo  {journal} {Phys.
  Rev. A}\ }\textbf {\bibinfo {volume} {90}},\ \bibinfo {pages} {025802}
  (\bibinfo {year} {2014})}\BibitemShut {NoStop}%
\bibitem [{Note2()}]{Note2}%
  \BibitemOpen
  \bibinfo {note} {Another consequence of (\ref {int}) is that the phase of the
  pump field is inherited by the squeezed vacuum created in mode $\protect
  \mathaccentV {hat}05E{a}$. Without loss of generality, we have set this phase
  to be 0.}\BibitemShut {Stop}%
\bibitem [{\citenamefont {{Demkowicz-Dobrzanski}}\ \emph
  {et~al.}(2014)\citenamefont {{Demkowicz-Dobrzanski}}, \citenamefont
  {{Jarzyna}},\ and\ \citenamefont {{Kolodynski}}}]{Raf2014}%
  \BibitemOpen
  \bibfield  {author} {\bibinfo {author} {\bibfnamefont {R.}~\bibnamefont
  {{Demkowicz-Dobrzanski}}}, \bibinfo {author} {\bibfnamefont {M.}~\bibnamefont
  {{Jarzyna}}}, \ and\ \bibinfo {author} {\bibfnamefont {J.}~\bibnamefont
  {{Kolodynski}}},\ }\href@noop {} {\bibfield  {journal} {\bibinfo  {journal}
  {ArXiv e-prints}\ } (\bibinfo {year} {2014})},\ \Eprint
  {http://arxiv.org/abs/1405.7703} {arXiv:1405.7703 [quant-ph]} \BibitemShut
  {NoStop}%
\bibitem [{\citenamefont {Herzog}\ \emph {et~al.}(1994)\citenamefont {Herzog},
  \citenamefont {Rarity}, \citenamefont {Weinfurter},\ and\ \citenamefont
  {Zeilinger}}]{PhysRevLett.72.629}%
  \BibitemOpen
  \bibfield  {author} {\bibinfo {author} {\bibfnamefont {T.~J.}\ \bibnamefont
  {Herzog}}, \bibinfo {author} {\bibfnamefont {J.~G.}\ \bibnamefont {Rarity}},
  \bibinfo {author} {\bibfnamefont {H.}~\bibnamefont {Weinfurter}}, \ and\
  \bibinfo {author} {\bibfnamefont {A.}~\bibnamefont {Zeilinger}},\ }\href
  {\doibase 10.1103/PhysRevLett.72.629} {\bibfield  {journal} {\bibinfo
  {journal} {Phys. Rev. Lett.}\ }\textbf {\bibinfo {volume} {72}},\ \bibinfo
  {pages} {629} (\bibinfo {year} {1994})}\BibitemShut {NoStop}%
\bibitem [{\citenamefont {Weinfurter}\ \emph {et~al.}(1995)\citenamefont
  {Weinfurter}, \citenamefont {Herzog}, \citenamefont {Kwiat}, \citenamefont
  {Rarity}, \citenamefont {Zeilinger},\ and\ \citenamefont
  {Zukowski}}]{NYAS:NYAS61}%
  \BibitemOpen
  \bibfield  {author} {\bibinfo {author} {\bibfnamefont {H.}~\bibnamefont
  {Weinfurter}}, \bibinfo {author} {\bibfnamefont {T.}~\bibnamefont {Herzog}},
  \bibinfo {author} {\bibfnamefont {P.~G.}\ \bibnamefont {Kwiat}}, \bibinfo
  {author} {\bibfnamefont {J.~G.}\ \bibnamefont {Rarity}}, \bibinfo {author}
  {\bibfnamefont {A.}~\bibnamefont {Zeilinger}}, \ and\ \bibinfo {author}
  {\bibfnamefont {M.}~\bibnamefont {Zukowski}},\ }\href {\doibase
  10.1111/j.1749-6632.1995.tb38956.x} {\bibfield  {journal} {\bibinfo
  {journal} {Annals of the New York Academy of Sciences}\ }\textbf {\bibinfo
  {volume} {755}},\ \bibinfo {pages} {61} (\bibinfo {year} {1995})}\BibitemShut
  {NoStop}%
\bibitem [{\citenamefont {Resch}\ \emph {et~al.}(2001)\citenamefont {Resch},
  \citenamefont {Lundeen},\ and\ \citenamefont
  {Steinberg}}]{PhysRevLett.87.123603}%
  \BibitemOpen
  \bibfield  {author} {\bibinfo {author} {\bibfnamefont {K.~J.}\ \bibnamefont
  {Resch}}, \bibinfo {author} {\bibfnamefont {J.~S.}\ \bibnamefont {Lundeen}},
  \ and\ \bibinfo {author} {\bibfnamefont {A.~M.}\ \bibnamefont {Steinberg}},\
  }\href {\doibase 10.1103/PhysRevLett.87.123603} {\bibfield  {journal}
  {\bibinfo  {journal} {Phys. Rev. Lett.}\ }\textbf {\bibinfo {volume} {87}},\
  \bibinfo {pages} {123603} (\bibinfo {year} {2001})}\BibitemShut {NoStop}%
\bibitem [{\citenamefont {Resch}\ \emph
  {et~al.}(2002{\natexlab{a}})\citenamefont {Resch}, \citenamefont {Lundeen},\
  and\ \citenamefont {Steinberg}}]{PhysRevLett.88.113601}%
  \BibitemOpen
  \bibfield  {author} {\bibinfo {author} {\bibfnamefont {K.~J.}\ \bibnamefont
  {Resch}}, \bibinfo {author} {\bibfnamefont {J.~S.}\ \bibnamefont {Lundeen}},
  \ and\ \bibinfo {author} {\bibfnamefont {A.~M.}\ \bibnamefont {Steinberg}},\
  }\href {\doibase 10.1103/PhysRevLett.88.113601} {\bibfield  {journal}
  {\bibinfo  {journal} {Phys. Rev. Lett.}\ }\textbf {\bibinfo {volume} {88}},\
  \bibinfo {pages} {113601} (\bibinfo {year} {2002}{\natexlab{a}})}\BibitemShut
  {NoStop}%
\bibitem [{\citenamefont {Resch}\ \emph
  {et~al.}(2002{\natexlab{b}})\citenamefont {Resch}, \citenamefont {Lundeen},\
  and\ \citenamefont {Steinberg}}]{PhysRevLett.89.037904}%
  \BibitemOpen
  \bibfield  {author} {\bibinfo {author} {\bibfnamefont {K.~J.}\ \bibnamefont
  {Resch}}, \bibinfo {author} {\bibfnamefont {J.~S.}\ \bibnamefont {Lundeen}},
  \ and\ \bibinfo {author} {\bibfnamefont {A.~M.}\ \bibnamefont {Steinberg}},\
  }\href {\doibase 10.1103/PhysRevLett.89.037904} {\bibfield  {journal}
  {\bibinfo  {journal} {Phys. Rev. Lett.}\ }\textbf {\bibinfo {volume} {89}},\
  \bibinfo {pages} {037904} (\bibinfo {year} {2002}{\natexlab{b}})}\BibitemShut
  {NoStop}%
\bibitem [{\citenamefont {Lundeen}\ and\ \citenamefont
  {Steinberg}(2009)}]{PhysRevLett.102.020404}%
  \BibitemOpen
  \bibfield  {author} {\bibinfo {author} {\bibfnamefont {J.~S.}\ \bibnamefont
  {Lundeen}}\ and\ \bibinfo {author} {\bibfnamefont {A.~M.}\ \bibnamefont
  {Steinberg}},\ }\href {\doibase 10.1103/PhysRevLett.102.020404} {\bibfield
  {journal} {\bibinfo  {journal} {Phys. Rev. Lett.}\ }\textbf {\bibinfo
  {volume} {102}},\ \bibinfo {pages} {020404} (\bibinfo {year}
  {2009})}\BibitemShut {NoStop}%
\bibitem [{\citenamefont {Hudelist}\ \emph {et~al.}(2014)\citenamefont
  {Hudelist}, \citenamefont {Kong}, \citenamefont {Liu}, \citenamefont {Jing},
  \citenamefont {Ou},\ and\ \citenamefont {Zhang}}]{Hudelist:2014dz}%
  \BibitemOpen
  \bibfield  {author} {\bibinfo {author} {\bibfnamefont {F.}~\bibnamefont
  {Hudelist}}, \bibinfo {author} {\bibfnamefont {J.}~\bibnamefont {Kong}},
  \bibinfo {author} {\bibfnamefont {C.}~\bibnamefont {Liu}}, \bibinfo {author}
  {\bibfnamefont {J.}~\bibnamefont {Jing}}, \bibinfo {author} {\bibfnamefont
  {Z.~Y.}\ \bibnamefont {Ou}}, \ and\ \bibinfo {author} {\bibfnamefont
  {W.}~\bibnamefont {Zhang}},\ }\href {http://dx.doi.org/10.1038/ncomms4049}
  {\bibfield  {journal} {\bibinfo  {journal} {Nat Commun}\ }\textbf {\bibinfo
  {volume} {5}} (\bibinfo {year} {2014})}\BibitemShut {NoStop}%
\bibitem [{\citenamefont {Ghobadi}\ \emph {et~al.}(2013)\citenamefont
  {Ghobadi}, \citenamefont {Lvovsky},\ and\ \citenamefont
  {Simon}}]{PhysRevLett.110.170406}%
  \BibitemOpen
  \bibfield  {author} {\bibinfo {author} {\bibfnamefont {R.}~\bibnamefont
  {Ghobadi}}, \bibinfo {author} {\bibfnamefont {A.}~\bibnamefont {Lvovsky}}, \
  and\ \bibinfo {author} {\bibfnamefont {C.}~\bibnamefont {Simon}},\ }\href
  {\doibase 10.1103/PhysRevLett.110.170406} {\bibfield  {journal} {\bibinfo
  {journal} {Phys. Rev. Lett.}\ }\textbf {\bibinfo {volume} {110}},\ \bibinfo
  {pages} {170406} (\bibinfo {year} {2013})}\BibitemShut {NoStop}%
\bibitem [{\citenamefont {Demkowicz-Dobrzanski}\ \emph
  {et~al.}(2012)\citenamefont {Demkowicz-Dobrzanski}, \citenamefont
  {Kolodynski},\ and\ \citenamefont {Guta}}]{Demkowicz2012}%
  \BibitemOpen
  \bibfield  {author} {\bibinfo {author} {\bibfnamefont {R.}~\bibnamefont
  {Demkowicz-Dobrzanski}}, \bibinfo {author} {\bibfnamefont {J.}~\bibnamefont
  {Kolodynski}}, \ and\ \bibinfo {author} {\bibfnamefont {M.}~\bibnamefont
  {Guta}},\ }\href {http://dx.doi.org/10.1038/ncomms2067} {\bibfield  {journal}
  {\bibinfo  {journal} {Nat Commun}\ }\textbf {\bibinfo {volume} {3}},\
  \bibinfo {pages} {1063} (\bibinfo {year} {2012})}\BibitemShut {NoStop}%
\bibitem [{\citenamefont {Knysh}\ \emph {et~al.}(2014)\citenamefont {Knysh},
  \citenamefont {Chen},\ and\ \citenamefont {Durkin}}]{knysh2014}%
  \BibitemOpen
  \bibfield  {author} {\bibinfo {author} {\bibfnamefont {S.~I.}\ \bibnamefont
  {Knysh}}, \bibinfo {author} {\bibfnamefont {E.~H.}\ \bibnamefont {Chen}}, \
  and\ \bibinfo {author} {\bibfnamefont {G.~A.}\ \bibnamefont {Durkin}},\
  }\href@noop {} {\bibfield  {journal} {\bibinfo  {journal} {arXiv preprint
  arXiv:1402.0495}\ } (\bibinfo {year} {2014})}\BibitemShut {NoStop}%
\bibitem [{\citenamefont {De~Martini}\ \emph {et~al.}(2008)\citenamefont
  {De~Martini}, \citenamefont {Sciarrino},\ and\ \citenamefont
  {Vitelli}}]{PhysRevLett.100.253601}%
  \BibitemOpen
  \bibfield  {author} {\bibinfo {author} {\bibfnamefont {F.}~\bibnamefont
  {De~Martini}}, \bibinfo {author} {\bibfnamefont {F.}~\bibnamefont
  {Sciarrino}}, \ and\ \bibinfo {author} {\bibfnamefont {C.}~\bibnamefont
  {Vitelli}},\ }\href {\doibase 10.1103/PhysRevLett.100.253601} {\bibfield
  {journal} {\bibinfo  {journal} {Phys. Rev. Lett.}\ }\textbf {\bibinfo
  {volume} {100}},\ \bibinfo {pages} {253601} (\bibinfo {year}
  {2008})}\BibitemShut {NoStop}%
\bibitem [{\citenamefont {Knott}\ \emph {et~al.}(2014)\citenamefont {Knott},
  \citenamefont {Proctor}, \citenamefont {Nemoto}, \citenamefont {Dunningham},\
  and\ \citenamefont {Munro}}]{knott}%
  \BibitemOpen
  \bibfield  {author} {\bibinfo {author} {\bibfnamefont {P.~A.}\ \bibnamefont
  {Knott}}, \bibinfo {author} {\bibfnamefont {T.~J.}\ \bibnamefont {Proctor}},
  \bibinfo {author} {\bibfnamefont {K.}~\bibnamefont {Nemoto}}, \bibinfo
  {author} {\bibfnamefont {J.~A.}\ \bibnamefont {Dunningham}}, \ and\ \bibinfo
  {author} {\bibfnamefont {W.~J.}\ \bibnamefont {Munro}},\ }\href@noop {}
  {\bibfield  {journal} {\bibinfo  {journal} {Phys. Rev. A}\ }\textbf {\bibinfo
  {volume} {90}},\ \bibinfo {pages} {033846} (\bibinfo {year}
  {2014})}\BibitemShut {NoStop}%
\bibitem [{Note3()}]{Note3}%
  \BibitemOpen
  \bibinfo {note} {In preparation}\BibitemShut {NoStop}%
\end{thebibliography}%

\end{document}